\documentclass[twoside,twocolumn,9pt]{article}
\usepackage{extsizes}
\usepackage[super,sort&compress,comma]{natbib} 
\usepackage[version=3]{mhchem}
\usepackage[left=1.5cm, right=1.5cm, top=1.785cm, bottom=2.0cm]{geometry}
\usepackage{balance}
\usepackage{mathptmx}
\usepackage{sectsty}
\usepackage{graphicx} 
\usepackage{lastpage}
\usepackage[format=plain,justification=justified,singlelinecheck=false,font={stretch=1.125,small,sf},labelfont=bf,labelsep=space]{caption}
\usepackage{float}
\usepackage{fancyhdr}
\usepackage{fnpos}
\usepackage[english]{babel}
\addto{\captionsenglish}{%
  
}
\usepackage{array}
\usepackage{droidsans}
\usepackage{charter}
\usepackage[T1]{fontenc}
\usepackage[usenames,dvipsnames]{xcolor}
\usepackage{setspace}
\usepackage[compact]{titlesec}
\usepackage{hyperref}
\definecolor{cream}{RGB}{222,217,201}

\usepackage{xr}

\makeatletter
\newcommand*{\addFileDependency}[1]{% argument=file name and extension
  \typeout{(#1)}
  \@addtofilelist{#1}
  \IfFileExists{#1}{}{\typeout{No file #1.}}
}
\makeatother

\newcommand{\qb}{\mathbf{q}}
\newcommand{\qq}{\mathbf{q}}

\newcommand{\rb}{\mathbf{r}}

\newcommand{\RR}{{\mathbf R}}
\newcommand{\rr}{{\mathbf r}}

\newcommand{\ee}{\varepsilon}

\newcommand{\dd}{{\,\text{d}}}

\usepackage{subcaption}
\usepackage{placeins}

\newcommand{\citen}[1]{%
  \begingroup
    \romannumeral-`\x % remove space at the beginning of \setcitestyle
    \setcitestyle{numbers}
    \cite{#1}
  \endgroup   
}

\newcommand{\mb}[1]{\mathbf{#1}}

\newcommand{\Sm}[1][]{%
   \ifthenelse{ \equal{#1}{} }
      { \Sigma^{-} }
      { \Sigma^{-}_{#1} }
}

\newcommand{\Sp}[1][]{%
   \ifthenelse{ \equal{#1}{} }
      { \Sigma^{+} }
      { \Sigma^{+}_{#1} }
}

\usepackage{cellspace}
\newcommand{\mg}[1]{$#1 \times #1 \times #1$}

\def\be{\begin{equation}}
\def\ee{\end{equation}}
\def\f{\frac}
\def\ni{\noindent}

\def\b{\beta}

\def\vare{\varepsilon}

\def\o{\omega}

\begin{document}

%\pagestyle{fancy}
%\thispagestyle{plain}
%\fancypagestyle{plain}{
%%%%HEADER%%%
%\renewcommand{\headrulewidth}{0pt}
%}
%%%END OF HEADER%%%

%%%PAGE SETUP - Please do not change any commands within this section%%%
\makeFNbottom
\makeatletter
\renewcommand\LARGE{\@setfontsize\LARGE{15pt}{17}}
\renewcommand\Large{\@setfontsize\Large{12pt}{14}}
\renewcommand\large{\@setfontsize\large{10pt}{12}}
\renewcommand\footnotesize{\@setfontsize\footnotesize{7pt}{10}}
\makeatother

\renewcommand{\thefootnote}{\fnsymbol{footnote}}
\renewcommand\footnoterule{\vspace*{1pt}% 
\color{cream}\hrule width 3.5in height 0.4pt \color{black}\vspace*{5pt}} 
\setcounter{secnumdepth}{5}

\makeatletter 
\renewcommand\@biblabel[1]{#1}            
\renewcommand\@makefntext[1]% 
{\noindent\makebox[0pt][r]{\@thefnmark\,}#1}
\makeatother 
\renewcommand{\figurename}{\small{Fig.}~}
\sectionfont{\sffamily\Large}
\subsectionfont{\normalsize}
\subsubsectionfont{\bf}
\setstretch{1.125} %In particular, please do not alter this line.
\setlength{\skip\footins}{0.8cm}
\setlength{\footnotesep}{0.25cm}
\setlength{\jot}{10pt}
\titlespacing*{\section}{0pt}{4pt}{4pt}
\titlespacing*{\subsection}{0pt}{15pt}{1pt}
%%%END OF PAGE SETUP%%%

%%%FOOTER%%%
\fancyfoot{}
\fancyfoot[LO,RE]{\vspace{-7.1pt}\includegraphics[height=9pt]{head_foot/LF}}
\fancyfoot[CO]{\vspace{-7.1pt}\hspace{11.9cm}\includegraphics{head_foot/RF}}
\fancyfoot[CE]{\vspace{-7.2pt}\hspace{-13.2cm}\includegraphics{head_foot/RF}}
\fancyfoot[RO]{\footnotesize{\sffamily{1--\pageref{LastPage} ~\textbar  \hspace{2pt}\thepage}}}
\fancyfoot[LE]{\footnotesize{\sffamily{\thepage~\textbar\hspace{4.65cm} 1--\pageref{LastPage}}}}
\fancyhead{}
\renewcommand{\headrulewidth}{0pt} 
\renewcommand{\footrulewidth}{0pt}
\setlength{\arrayrulewidth}{1pt}
\setlength{\columnsep}{6.5mm}
\setlength\bibsep{1pt}
%%%END OF FOOTER%%%

%%%FigURE SETUP - please do not change any commands within this section%%%
\makeatletter 
\newlength{\figrulesep} 
\setlength{\figrulesep}{0.5\textfloatsep} 

\newcommand{\topfigrule}{\vspace*{-1pt}% 
\noindent{\color{cream}\rule[-\figrulesep]{\columnwidth}{1.5pt}} }

\newcommand{\botfigrule}{\vspace*{-2pt}% 
\noindent{\color{cream}\rule[\figrulesep]{\columnwidth}{1.5pt}} }

\newcommand{\dblfigrule}{\vspace*{-1pt}% 
\noindent{\color{cream}\rule[-\figrulesep]{\textwidth}{1.5pt}} }

\makeatother
%%%END OF FIGURE SETUP%%%

%%%TITLE, AUTHORS AND ABSTRACT%%%
\twocolumn[
  \begin{@twocolumnfalse}
%{\includegraphics[height=30pt]{head_foot/PCCP}\hfill\raisebox{0pt}[0pt][0pt]{\includegraphics[height=55pt]{head_foot/RSC_LOGO_CMYK}}\\[1ex]
%\includegraphics[width=18.5cm]{head_foot/header_bar}}\par
\vspace{1em}
%\sffamily
%\begin{tabular}{m{4.5cm} p{13.5cm} }
{\centering
\noindent\LARGE{\textbf{
%Nonpolarons: Polarons in non-polar systems
Spectroscopic signatures of nonpolarons : the case of diamond}} % \\
\vspace{0.3cm}% & \vspace{0.3cm} \\

 \noindent\large{Joao C. de Abreu,\textit{$^{1}$} Jean Paul Nery,\textit{$^{2}$} Matteo Giantomassi,\textit{$^{3}$} Xavier Gonze\textit{$^{3,4}$} and Matthieu J. Verstraete\textit{$^{1}$}}% \\

\vspace{1 em}

\textit{$^{1}$~nanomat/Q-MAT/CESAM and European Theoretical Spectroscopy Facility, Universit\'e de Li\`ege, B-4000 Belgium}

\textit{$^{2}$~Dipartimento di Fisica, Universit\`a di Roma La Sapienza, I-00185 Roma, Italy }

\textit{$^{3}$~UCLouvain, Institute of Condensed Matter and Nanosciences (IMCN), Chemin des \'Etoiles~8, B-1348 Louvain-la-Neuve, Belgium }

\textit{$^{4}$~Skolkovo Institute of Science and Technology, Moscow, Russia }
\vspace{2 em}

\noindent\normalsize{Polarons are quasi-particles made from electrons interacting with vibrations in crystal lattices. They derive their name from the strong electron-vibration polar interaction in ionic systems, that induces associated spectroscopic and optical signatures of such quasi-particles in these materials. In this paper, we focus on diamond, a non-polar crystal with inversion symmetry which nevertheless shows characteristic signatures of polarons, better denoted ``nonpolarons'' in this case. The polaronic effects are produced by short-range crystal fields with only a small influence of  long-range quadrupoles. The many-body spectral function has a characteristic energy dependence, showing a plateau structure that is similar to but distinct from the satellites observed in the polar Fr\"ohlich case. 
The temperature-dependent spectral function of diamond is determined by two methods: the standard Dyson-Migdal approach, which calculates electron-phonon interactions within the lowest-order expansion of the self-energy, and the cumulant expansion, which includes higher orders of electron-phonon interactions. The latter corrects the nonpolaron energies and broadening, providing a more realistic spectral function, which we examine in detail for both conduction and valence band edges.}
}
%\end{tabular}

 \end{@twocolumnfalse} \vspace{1.0cm}

  ]
%%%END OF TITLE, AUTHORS AND ABSTRACT%%%

%%%FONT SETUP - please do not change any commands within this section
\renewcommand*\rmdefault{bch}\normalfont\upshape
\rmfamily
\section*{}
\vspace{-1cm}

%%%FOOTNOTES%%%

%\footnotetext{\dag~Electronic Supplementary Information (ESI) available: [details of any supplementary information available should be included here]. See DOI: 10.1039/cXCP00000x/}

%%%END OF FOOTNOTES%%%

\raggedbottom

\section{Introduction}

Interactions between electrons and vibrational modes of solids (phonons) create composite bound states known as polarons. Most of the attention in the field has quite naturally been focused in systems where these effects are expected to be strong, e.g. polar materials \citep{Meevasana2010}, their vacancies \citep{Janotti2014}, molecular crystals\citep{Chaikin1972} or 2D-materials \citep{McKenna2012}. In these systems the electrons interact among others with long-range dipole fields induced by displaced ions. Although covalent materials have no dipole moments, one could expect long-range (LR) quadrupole fields to contribute to the formation of polarons, as inferred from their significant impact on the carrier mobility in Si \citep{Brunin2020,Park2020}.

Some covalent systems with strong electron-phonon (e-ph) interactions show conductivity induced by hopping of small polarons, e.g.: disordered systems such as chalcogenide glasses \citep{Emin1982}, molecular crystals such as S$_8$ \citep{Gibbons1966}, rare gas solids such as Xe \citep{Howe1971}, or 2D phosphorene and arsenene\citep{Vasilchenko2021}, where the polaron is localized in lone-pair orbitals. 
Large polarons in covalent materials have been historically neglected, and were dubbed ``nonpolarons'' by Emin \citep{Emin1994}. 
We note that polaronic signatures were found in doped diamond with hydrogen-terminated surface having a negative electron affinity \citep{Rameau2011}.
In the present paper we study polaronic effects in intrinsic diamond, to quantify from first principles the binding and spectral signatures of polarons in non-polar materials.

The detection of polarons in a crystal often relies on angle-resolved photoemission spectroscopy (ARPES), which measures the kinetic energy and angular distribution of electrons excited by incident light. 
These quantities are directly related to the number of states available, as a function of energy and momentum. Signatures of polarons in ARPES experiments can be found in cuprate superconductors\citep{Shen2004}, ionic 3D\citep{Mohamed2019,Chen2015,Emori2014,Moser2013} and 2D\citep{Chen2015,Kang2018} crystals, ferromagnetic materials\citep{Riley2018}, and interfaces\citep{Cancellieri2018}. 
In the simplest model, neglecting surface effects, ARPES can be related to the one-electron spectral function, which is the central property of interest here.

Besides (non)polaron binding and the renormalization of the direct electronic band gap\citep{Antonius2014}, the effects of phonons in non-polar materials can also be seen in the optical excitation of carriers in indirect band gap semiconductors\citep{Lautenschlager1986}. The scattering of carriers by phonons dominates transport mechanisms at high temperature, hence 
an appropriate description of the spectral function is essential to
calculate transport properties
including many-body effects.
%MG: The semi-classical Boltzmann equation does not require A(w).
Important experimental observables are the thermoelectric conductivity \citep{Tong2019}, the charge carrier mobility/conductivity\citep{Ziman1960}, or superconductivity\citep{Marsiglio2008}.
    
First-principles calculations of the spectral function at the valence band maximum (VBM) and conduction band minimum (CBM) for (polar) LiF and MgO were examined by Nery \textit{et al} \citep{Nery2018}  within a zero temperature formalism. In this paper, we expand their approach by studying a non-polar material, diamond, and including finite temperature effects. We will focus on the renormalization of electronic energies at T=0 K (the zero point renormalization, or ZPR), their temperature dependence, and the emergence of nonpolaronic signatures. %Calculations of the ground-state 
Vibrational properties and e-ph matrix elements
are obtained using density functional perturbation theory (DFPT)\citep{Gonze1997,Baroni2001} 
%and the inclusion of the 
while the interaction between electrons and phonons, which leads to the formation of a quasi-particle (QP), is treated using many-body perturbation theory (MBPT)\citep{Martin2016}.

The article is organized as follows.
Section \ref{sec:Methods} summarizes the most important theoretical aspects of the e-ph problem with particular emphasis on the different approaches that can be used 
to compute spectral functions and QP energies.
More specifically, we compare the standard Dyson equation in the Migdal approximation (DM)\citep{Dyson1949,Migdal1958} with the cumulant-expansion (CE) method \citep{Kubo1962,Gunnarsson1994,Kas2014}, which includes higher order diagrams in the self-energy. 
The CE was previously shown to provide accurate results for e-e interactions\citep{Aryasetiawan1996} and also to improve plasmonic polaron satellite energies\citep{Guzzo2012,Caruso2015}.
In Section \ref{sec:Results} we describe spectral signatures in diamond and show that the CE improves with respect to DM.
In addition, we show that LR quadrupole fields do not contribute strongly to the spectral signals; the latter are mostly created by local crystal fields. 
We also include an Electronic Supplementary Information (ESI) which analyses and clarifies important aspects of numerical convergence, analytical transformations for the cumulant expansion, and finite temperature effects.
Atomic units are used everywhere unless explicitly noted.

\section{Methods}\label{sec:Methods}

We study the interaction between bare electrons and bare phonons by assuming that it can be treated with MBPT techniques. It is important to note that for polarons this is not always possible. %assumption is not always guaranteed. 
In strongly interacting cases small localized polarons can be formed, which cannot be treated perturbatively. In ESI \ref{subsec: Electronic Phonon} we summarize the methods used for the ground-state calculations.

\subsection{Self-energy}\label{subsection:Selfenergy}
Following Ref.~\citen{Nery2018}
we refer to the lowest order e-ph self-energy (second order in the atomic displacements) as the Fan-Migdal (FM) self-energy\citep{Migdal1958}. It includes two terms,
the static Debye-Waller (DW)\citep{Antoncik1955} and the dynamic Fan\citep{Fan1951} term,
\begin{equation}
       \Sigma_{n \mb{k} } ( \omega) = \Sigma^{\text{DW}}_{n \mb{k}} + \Sigma^{\text{Fan}}_{n \mb{k}} ( \omega ) .
\label{eq:FanDW}
\end{equation}
where
    \begin{equation}\label{eq:DW}
      \Sigma^{\text{DW}}_{n \mb{k}} = \f{1}{N_\mathbf{q}} \sum_{j \mb{q}} g^{\mathrm{DW},j\mathbf{q},j-\mathbf{q}}_{nn\mathbf{k}}  \left( 2n_{j \mb{q}} (T) + 1 \right)
      % \Sigma^{\text{DW}}_{n \mb{k}} = -\f{1}{N_\mathbf{q}}\sum_{j \mb{q}}\sum_{m} \frac{\left| g_{n m \mb{k}}^{\text{DW},j \mb{q}=0} \right|^2}{\varepsilon_{n k} - \varepsilon_{m \mb{k}} } \left( n_{j \mb{q}} (T) + \frac{1}{2} \right)
   \end{equation}
   and
%
    %\begin{widetext}
        \begin{equation}
        \begin{split}
            & \Sigma^{\text{FAN}}_{n \mb{k}} ( \omega ) = \frac{1}{N_\mathbf{q}} \sum_{m} \sum_{j \mb{q} } \left| g_{n m \mathbf{k}}^{j \mb{q}} \right|^2 \times \\
            & \times \left( \underbrace{\frac{n_{j \mb{q}} (T) + f_{m \mb{k}+\mb{q}}(T)}{\omega - \varepsilon_{m \mb{k}+\mb{q}} + \omega_\mb{j q} + i \eta}}_{\Sm} + \underbrace{\frac{n_{j \mb{q}} (T) +1 - f_{m \mb{k}+\mb{q}}(T)}{\omega - \varepsilon_{m \mb{k}+\mb{q}} - \omega_\mb{j q} + i \eta}}_{\Sp} \right).
        \end{split}            
            \label{eq:Fan}
        \end{equation}
   % \end{widetext}
%
with $\eta$ a positive infinitesimal (retarded self-energy)~\citep{Giustino2017}.
In the equations above, we use
$\varepsilon_{m \mathbf{k}}$ to denote the energy of the electronic state
with band index $m$, wavevector $\mb{k}$ and Fermi-Dirac occupation number $f_{m \mb{k}}$.
The phonon frequencies are denoted with $\omega_{j \mb{q}}$ with $j$ the mode index, $\mb{q}$ the phonon wave vector, and $n_{j \mb{q}}$ the Bose-Einstein occupation number. 
Finally, the symbol $g_{nm \mb{k}}^{j \mb{q}}$ denotes the first order e-ph matrix element. 

In principle, the DW matrix element $g^{\text{DW}}$ should be computed as the second-order derivative of the self-consistent potential with respect to the phonon displacement.
In practice, we use the acoustic sum rule, together with the rigid-ion approximation\citep{Allen1976}, to 
express $g^{\text{DW}}$ in terms of the first order e-ph matrix elements\citep{Ponce2015,Giustino2017}.
Inside the parentheses of Eq.~\ref{eq:Fan}, there are two terms with the Bose Einstein factors $n$ and $n+1$, which we label $\Sm$ and $\Sp$. Each term represents in turn two separate scattering events, into and out of state $n\mb{k}$, which correspond to absorption and emission of phonons, pairing with the appropriate electron or hole state.

\subsection{Dyson-Migdal approach}\label{subsection:DysonMigdal}

The interacting Green's function $G$ can be expressed 
in terms of the initial bare propagator $G^{0}$
and the exact self-energy via the Dyson equation\citep{Mahan2000}.
In the diagonal approximation\cite{Antonius2015},
the off-diagonal matrix-elements of $\Sigma$ in the Bloch basis set are assumed to be negligible, and the Dyson equation reduces to
    \begin{equation} \label{eq: Dyson equation}
       G_{n\mb{k}} (\omega) = \frac{1}{\left(G^{0}_{n\mb{k}} (\omega) \right)^{-1} - \Sigma_{n\mb{k}} (\omega)}.
    \end{equation}

Finally, the spectral function $A$
is given by 
\begin{equation}\label{eq:AbsorptionSpectrum}
    A_{n \mb{k} } (\omega) = - \frac{1}{\pi} \Im m G^R_{n \mb{k} } (\omega).
\end{equation}   
where $G^R$ is the retarded Green's function.
From Eq.~\ref{eq: Dyson equation} and Eq.~\ref{eq:AbsorptionSpectrum}, one obtains~\citep{Giustino2017}
\begin{equation}\label{eq:Dyson_A}
 A_{n \mb{k}}(\omega) = -\frac{1}{\pi} 
\frac{\Im \Sigma_{n \mb{k} }(\omega)}{(\omega - \varepsilon_{n \mb{k}} - \Re \Sigma_{n \mb{k}}(\omega))^2 + \Im \Sigma_{n \mb{k}}(\omega)^2}
\end{equation}   
At this point, it is worth stressing that in practical applications it is customary to evaluate Eq.\ref{eq:Dyson_A} using the lowest order FM self-energy (Eq. \ref{eq:FanDW})
evaluated with bare electron/phonon quantities (one-shot method).
This approach, which neglects self-consistency effects and vertex corrections\citep{Nery2018,Giustino2017},
will be referred to as the Dyson-Migdal (DM) approximation in what follows.
According to previous studies in polar materials\citep{Nery2018},
the DM approach usually yields poor QP energies and spectral weights when compared with high-quality Monte Carlo calculations for the Fr\"{o}hlich model\citep{Mishchenko2000}.
Moreover, the position of the DM satellite relative to the QP peak is often inaccurate and far from the expected value,
which should match the phonon frequency $\omega_{LO}$ of the LO 
mode. 
A promising route for going beyond the DM approximation is the cumulant expansion detailed in the next section.
\subsection{Cumulant expansion}\label{subsection:Cumulant expansion}

The Green's function in the time domain can be rewritten in an exponential form (cumulant expansion)
using Kubo's formula\cite{Kubo1962}
   \begin{equation} \label{eq:GreenFunctionCumulant}
       G_{n \mb{k} } (t) = G^0_{n \mb{k} } (t) e^{C_{n \mb{k} } (t)},
   \end{equation}
where
   \begin{equation} \label{eq:CumulantSeries}
       C_{n \mb{k} } (t) = \sum_{i=2}^{\infty} C_{n \mb{k} } ^{(i)} (t)
   \end{equation}
is the sum of the cumulants of the i-th order. The cumulant function in Eq.~(\ref{eq:CumulantSeries}) can achieve accurate results\citep{Gumhalter2005} for our problem using just $i=2$, and is exact for a (fully localized) core electron interacts with a phonon\citep{Langreth1970}. 
   
The cumulant functions can be determined by expanding the exponential in Eq. (\ref{eq:GreenFunctionCumulant}) and comparing powers with the standard Feynman expansion of the Green's function\citep{Aryasetiawan1996}. 
Using the Fan-Migdal self-energy (Fan and DW terms), one obtains
   \begin{equation} 
       C_{n \mb{k} } (t) = \int d \omega \beta_{n \mb{k} }  (\omega) \frac{e^{-i \omega t} + i \omega t - 1}{\omega^2}
       \label{eq:Cumulant}
   \end{equation}
where
   \begin{equation}
       \beta_{n \mb{k} }  (\omega) = \frac{1}{\pi} \left| \Im m \Sigma_{n \mb{k} }^{\text{Fan}} (\omega + \varepsilon_{n \mb{k} }) \right|
   \end{equation}
while the DW self-energy appears as a pure shift of the QP energy,
   \begin{equation} \label{eq: GreenFunctionTimeDomain}
       G_{n \mb{k} } (t) = -i \theta(t) e^{-i \left( \varepsilon_{n \mb{k} } + \Sigma^{\text{DW}}_{n \mb{k} } \right) t} e^{C_{n \mb{k} } (t)}.
   \end{equation}
The three terms in Eq. (\ref{eq:Cumulant}) have different effects on the spectral function. The first one gives rise to satellites, the second term shifts the QP peak, while the third term corrects the QP weight. The second term is calculated using the Kramers-Kronig relations. Further details can be found in section \ref{subsec:Kramers-Kronig} of the ESI.

The spectral function is obtained by applying the inverse Fourier transform to the Green's Function in the time domain, Eq. (\ref{eq: GreenFunctionTimeDomain}), and inserting it into Eq. (\ref{eq:AbsorptionSpectrum}). It can be shown that the CE Green's function is exact to second order, and all higher order terms are included, though in an approximate way, while DM only includes the exact second order terms\cite{Guzzo2011,Hedin1980}.
In the long time limit, the cumulant has an affine asymptotic behaviour which contains its contributions to the lifetime and lineshape of the QP state,
\begin{equation} \label{eq:GreenFunctionInftyLimit}
        \begin{split}
            G_{n \mb{k} } (t \to \infty) \approx G^{0}_{n \mb{k} } (t) \exp \left[ - w_{k} + \left(- \Gamma_{n \mb{k} } + i \Lambda_{n \mb{k} } \right) t  \right]
        \end{split}
\end{equation}
where $w_k$ is a constant, $\Gamma_{n \mb{k} } = \left| \Im m \Sigma_{ n \mb{k}}( \omega = \varepsilon^{0}_{n \mb{k} }) \right|$ is the decay rate, $\Lambda_{n \mb{k} } = \Re e  \Sigma_{n \mb{k}}(\omega = \varepsilon^{0}_{n \mb{k} })$, and $G^{0}$ depends on $t$ through an exponential with the bare electron energy.

\subsection{Energy renormalization}\label{subsec: Energy renormalization}
   
In this section we summarize three commonly used approximations to compute the QP energy from the e-ph self-energy.
If we ignore the frequency dependence of the imaginary part of the self-energy, 
one obtains that the main peak of the DM spectral function (Eq.~\ref{eq:Dyson_A})
is located at the energy $\varepsilon^{\text{NL}}_{n \mb{k}}$ 
that solves the \emph{non-linear} (NL) QP equation
\begin{equation}\label{eq:SCQPenergy}
        \varepsilon^{\text{NL}}_{n \mb{k} } = \varepsilon^{0}_{n \mb{k} } + \Re e \Sigma_{n \mb{k}}(\omega = \varepsilon^{\text{NL}}_{n \mb{k} } ).
\end{equation}
This equation must be solved numerically using \emph{e.g.} root-finding algorithms
that require the knowledge of $\Sigma_{n \mb{k}}(\omega)$
for several frequencies.
The problem can be significantly simplified if we 
assume the QP correction 
$\varepsilon^{\text{NL}}_{n \mb{k} } - \varepsilon^{0}_{n \mb{k} }$ to be small.
In this case one can 
expand the self-energy to linear order around the KS energy, $\varepsilon^0_{n \mb{k}}$ to obtain the \emph{linearized} QP equation
\begin{equation}\label{eq:linearQPenergy}
        \varepsilon^{\text{linear}}_{n \mb{k} } = \varepsilon^{0}_{n \mb{k} } + Z_{n \mb{k} } \Re e \Sigma_{n \mb{k}}(\omega = \varepsilon^{0}_{n \mb{k} } )
\end{equation}
with the renormalization factor $ Z_{n \mb{k} }$ given by
\begin{equation}
         Z_{n \mb{k} } = \left( 1 - \Re e \left. \frac{\partial \Sigma_{n \mb{k} } (\omega, T)}{\partial \omega} \right|_{\omega = \varepsilon^{0}_{n \mb{k} }} \right)^{-1},
\end{equation}
that is approximately equal to the area under the QP peak of the spectral function.
Finally, in Rayleigh-Schr\"{o}dinger perturbation theory, also known as the on-the-mass-shell (OMS) approach, the energy correction is just given by the self-energy evaluated at the KS energy~\citep{Giustino2017}
\begin{equation}\label{eq:MSQPenergy}
        \varepsilon^{\text{OMS}}_{n \mb{k} } = \varepsilon^{0}_{n \mb{k} } + \Re e \Sigma_{n \mb{k}}(\omega = \varepsilon^{0}_{n \mb{k} } ).
\end{equation}

The evaluation of the energy correction is also carried out for the CE approach. The effective CE self-energy is determined by inverting Eq. (\ref{eq: Dyson equation}) after inserting the CE and non-interacting Green's functions. Then the NL equation is used to obtain eigenenergies. Further details over the effective CE self-energy are described in ESI \ref{sec:SelfEnergy}.

In section \ref{subsec:SpectralFunctionARPESTemperature}, we will check the results of the different approaches for the band edges of diamond, showing that the OMS approximation using DM produces QP shifts very close to the CE (as already observed for polar materials\citep{Nery2018}).

\subsection{Long-range and short-range potentials}\label{subsec:QuadrupolesMethods}

Converging e-ph calculations requires very dense $\mb{q}$-grids that are prohibitively expensive for DFPT.
For this reason, we employ the Fourier-based interpolation scheme initially proposed by Eiguren\citep{Eiguren2008} to interpolate the e-ph scattering potentials on arbitrarily dense $\mb{q}$-meshes.
The potential interpolation scheme has an important advantage for the e-ph matrix elements of high unoccupied states, which are very delicate to access accurately with Wannier Function based interpolation methods\cite{Giustino2007}.

The e-ph scattering potential presents a non-analytical behaviour~\citep{Vogl1976}
in the long wave-length limit ($\mb{q} \to 0$)
associated to a LR behaviour in real-space
which requires a specialized numerical treatment.
Over the past decade, this problem has been subject to several investigations that lead to a well established procedure:
The dipole fields for $\mb{q} \to 0$ are treated using a Fr\"{o}hlich-like potential\citep{Verdi2015,Sjakste2015} which depends on the Born effective charges and diverges as $1/q$. Recently\citep{Brunin2020,Brunin2020_2,Jhalani2020}, the treatment of LR contributions has been generalized to include contributions generated by dynamical quadrupoles~\citep{Royo2019}.
As the Born effective charges of diamond are zero, in what follows we focus on the treatment of the quadrupole interaction.

In non-polar materials, the Fourier interpolation of the e-ph potentials proceeds by first removing the non-analytical long-range contribution induced by the displacement of the 
$\kappa$-th atom along the Cartesian direction $\alpha$
using the generalized quadrupole model 
\begin{equation} \label{eq:VooglPotential}
\begin{split}
    V^{\mathcal{L}}_{\kappa \alpha, \mb{q}} ( \mb{r}) =& \frac{4 \pi}{\Omega} \sum_{\mb{G} \neq - \mb{q} } \frac{( q_{\beta} + G_{\beta}) ( q_{\gamma} + G_{\gamma}) }{(q_{\delta} + G_{\delta} ) \epsilon^{\infty}_{\delta \delta'} ( q_{\delta'} + G_{\delta'})} \\
    &\times \frac{1}{2} Q^{\beta \gamma}_{\kappa \alpha} e^{i (q_{\eta} + G_{\eta} ) ( r_{\eta} - \tau_{\kappa \eta})} e^{- \frac{ \left| \mb{q} + \mb{G} \right|^2 } { 4 \alpha } },
\end{split}
\end{equation}
where $\Omega$ is the unit cell volume, $Q$ is the dynamical quadrupole tensor, $\mathbf{G}$ are the reciprocal lattice vectors, $\mathbf{\tau}_\mathbf{\kappa}$ is the position of the atom %$\kappa$ 
in the unit cell, and $\epsilon^{\infty}$ is the electronic dielectric tensor in Cartesian coordinates
(repeated indices are implicitly summed over).
The last exponential term, $ e^{- \frac{ \left| \mb{q} + \mb{G} \right|^2 } { 4 \alpha } } $, is a Gaussian filter\citep{Brunin2020_2} with a variance of $\sqrt{\alpha}$. 
The resulting short-range (SR) potentials, which are smooth and analytic in $\mb{q}$ space, are then used to build the scattering potential $W_{\kappa,\alpha}(\rr, \RR)$ in the real-space supercell
(see \emph{e.g} Eq 12 in Brunin \textit{et al}\citep{Brunin2020_2}).
The short-range $W$ is Fourier-interpolated on a much denser $\mb{q}$-grid 
and the non-analytic LR terms 
are finally added back to get the total scattering potential. .
Interestingly, one can use this interpolation technique to compute e-ph matrix elements in which only the LR (SR) part of the scattering potential is included.
In section \ref{subsec:Quadrupolemodel}, we will analyze the separate contributions from SR and LR potential and the effect on self energies and spectral functions.

\section{Results}\label{sec:Results}

All calculations in this work are performed using the ABINIT\citep{Gonze2020,Romero2020} software. Many-body calculations are presented in the following section.
Further details on ground-state and DFPT calculations
can be found in ESI \ref{subsecapp:Computational details}.

\subsection{Zero-point renormalization}

\begin{figure}[h]
    \includegraphics[width=\columnwidth]{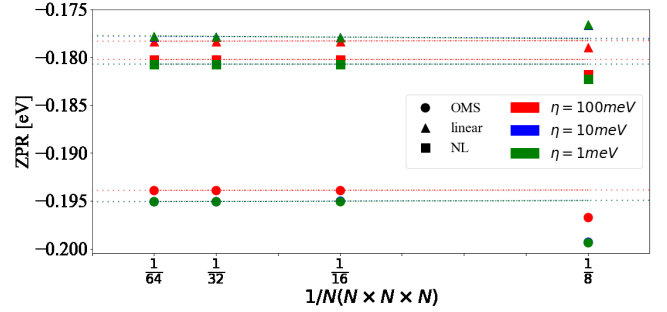}
    \caption{ZPR for the CBM as a function of the number of divisions in the $\mathbf{q}$-mesh and different values of $\eta$.
    The data was fitted with a linear model and extrapolated for $1/N \rightarrow 0$. Calculations were done with the OMS approach Eq.(\ref{eq:MSQPenergy}) (circles), the linear approximation Eq.(\ref{eq:linearQPenergy}) (triangles) and the non-linear approach Eq.~(\ref{eq:SCQPenergy}) (squares). 
    The values obtained with $\eta=10$ meV and $\eta=1$ meV
    are 
    indistinguishable on the scale of the graph.
    }
    \label{fig:ZPR_Diamond}
\end{figure}

Accurate ZPR calculations require a careful convergence study with respect to the BZ sampling and the finite value of $\eta$. 
Figure \ref{fig:ZPR_Diamond} shows the results of such a convergence study 
for the CBM of diamond. 
%with respect to the phonon wave-vector sampling and the value of $\eta$ 
A uniform $16 \times 16 \times 16$ 
$\mb{q}$-grid sampling with $\eta = 10$ meV gives converged values for the ZPR, but we will see other quantities are more sensitive. 
Throughout the paper, we will refer to a $\mb{q}$-grid with size $N \times N \times N $ using just $N$.
The ZPR for the VBM converges with the same parameters as the CBM, giving a total ZPR for the band gap of $-0.325$ eV within the OMS equation.
The ZPRs for the VBM, CBM and the band gap obtained with the
three approximations discussed 
in Section~\ref{subsec: Energy renormalization}
are summarized in Table~\ref{tab:ZPR_CBM_VBM_BG}. 
At this level of theory, we expect OMS to provide the most accurate results by analogy with polar materials, where OMS is in good agreement  with high-quality Monte Carlo methods\citep{Nery2018}.
This might occur due to a fortuitous cancellation between the errors coming from the lack of higher order e-ph interactions, and the evaluation of the self-energy at the KS energy, a non-self-consistent calculation of the QP energy.

\begin{table}[h]
 \begin{tabular*}{0.48\textwidth}{@{\extracolsep{\fill}}c|ccc}
& CBM & VBM & Band gap \\ \hline
OMS & -0.196 & 0.130 & -0.325 \\
linear & -0.177 & 0.118 & -0.295 \\ 
NL & -0.180 & 0.119 & -0.299 \\
 \end{tabular*}
 \caption{ Converged ZPR values [eV] for CBM, VBM and band gap at $T=0K$ using the OMS approach Eq.(\ref{eq:MSQPenergy}), the linear approximation Eq.(\ref{eq:linearQPenergy})) and the non-linear approach Eq.~(\ref{eq:SCQPenergy}).}
 \label{tab:ZPR_CBM_VBM_BG}
\end{table}
%\end{center}

%\endgroup

The ZPR for polar materials, such as LiF and MgO, are largely dominated by the Fr\"{o}hlich interaction\citep{Nery2018}. 
In diamond, there are no dipole contributions, yet the ZPR is similar to that of LiF and MgO in relative terms. 
The Fr\"{o}hlich ZPR of MgO \citep{Nery2018,Roessler1967} and LiF \citep{Nery2018,Roessler1967_2} for the CBM is approximately 4\% and 1\%, respectively, of the experimental band gap. Similarly, the ZPR of diamond \citep{Madelung2004} is about 4\%. 
This shows that accurate computations of band gaps require the inclusion of e-ph interaction even in homo-polar crystals.
It should be noted, however, that electron-electron interactions beyond the KS-DFT mean-field approximation are absent in our calculations. These corrections may vary depending on the wave-vector, e.g., for the
indirect band gap of diamond the GW correction is around 0.02 eV while  one obtains  about 0.2 eV 
for the direct band gap\citep{Karsai2018,Antonius2014}.
We also ignored thermal expansion, zero-point lattice expansion~\cite{Miglio2020,Brousseau2022} and further anharmonic effects. Other calculations including these phenomena are detailed in Table \ref{tab:zpr}.

\subsection{Interplay between $\Im m \Sigma$ and $\eta$ }

In this section, we analyze the convergence of the imaginary part of the e-ph self-energy with respect to the $\bf{q}$-sampling and the broadening parameter $\eta$.
This convergence study is needed because the CE is rather sensitive to the quality of the input FM self-energy as detailed in the next section. 
According to our numerical tests, indeed,
the real part of the self-energy at the KS energy %for $T \to 0$ K 
converges with relatively coarse $\mb{q}$-meshes and large $\eta$ provided that enough empty states are included in the calculation.
On the contrary, the imaginay part
requires much denser $\mb{q}$-grids and smaller $\eta$. 
To elucidate this point,
we compare finite-$\eta$ results with those
obtained with the more accurate linear tetrahedron method\cite{Blochl1994} 
that is considered as a reference value.

\begin{figure}[h]
    \includegraphics[width=\columnwidth]{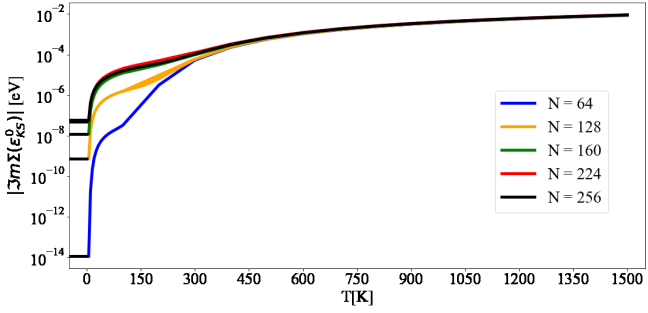}
    \caption{Convergence with $N$ of the imaginary part of the self-energy at the CBM evaluated at the KS energy as a function of temperature. Calculations are done with the DM approach and the tetrahedron method. 
    The lowest T considered is 5 K.
    Each curve has been extended to negative T to better visualize the convergence. The y-axis is in logarithmic scale.}
    \label{fig:ImSigmaTetrahedron}
\end{figure}

\begin{figure}[h]
    \centering
    \includegraphics[width=\columnwidth]{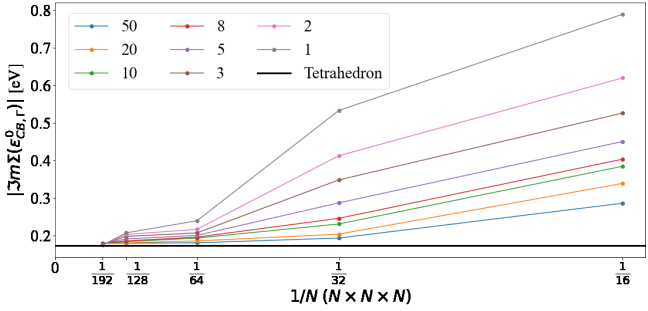}
    \caption{ Convergence of $\Im m \Sigma_{c \Gamma} ( \varepsilon^0_{c \Gamma})$ using the DM approach with $N$ for different values of $\eta= 50, 20, 10, 8, 5, 3, 2$, and $1$ meV (indicated in the legend) at 300 K. The black line denotes the reference value obtained with the tetrahedron method and $N=160$.}
    \label{fig:ImSigmaConvCBGamma}
\end{figure}

\begin{figure}[htb]
    \centering
    \includegraphics[width=\columnwidth]{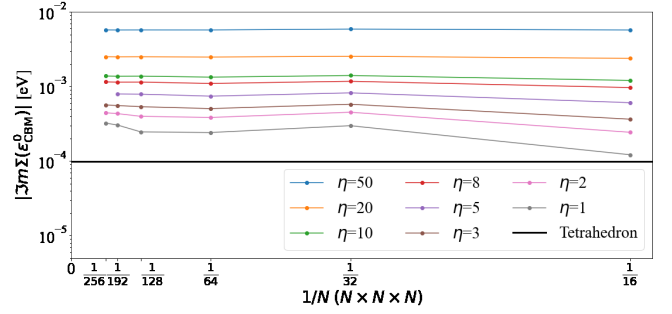}
    \caption{Convergence of $\Im m \Sigma _{\text{CBM}}(\varepsilon^0_{\text{CBM}})$ using the DM approach with $N$, for different values of $\eta$, at 300  K. Exactly at the CBM, the standard method converges slowly down to the tetrahedron value, as the value of $\Sigma$ is very small and comparable to the ``infinitesimal'' $\eta$ itself. Tetrahedron is calculated with $N=160$. The y-axis is in logarithmic scale.}
    \label{fig:ImSigmaConvCBM}
\end{figure}

Figure~\ref{fig:ImSigmaTetrahedron} shows the convergence of the imaginary part of $\Sigma$ at the CBM using the tetrahedron scheme. 
Above 5 K, $\Im m \Sigma(\varepsilon^0_{\text{CBM}})$ converges at $N=160$, and from 5 K to 60 K, there is a steep increase of $\Im m \Sigma(\varepsilon^0_{\text{CBM}})$ from $6\times10^{-8}$ eV to $10^{-5}$ eV, respectively. At 300 K, $\Im m \Sigma_{\text{CBM}}(\varepsilon^0_{\text{CBM}})$ reaches $10^{-4}$ eV and at 1500 K it is almost $10^{-2}$ eV.
In Figs.~\ref{fig:ImSigmaConvCBGamma} and \ref{fig:ImSigmaConvCBM}, we finally compare finite $\eta$ results with the converged tetrahedron values.
Figure~\ref{fig:ImSigmaConvCBGamma} shows that
$\Im m \Sigma_{c \Gamma} (\varepsilon^0_{c \Gamma})$, where $c \Gamma$ means bottom of the conduction band at $\Gamma$, 
converges towards the tetrahedron value 
for large $N$ and small $\eta$.
Using $N=128$, $\Im m \Sigma_{c \Gamma} (\varepsilon^0_{c \Gamma})$ differs from the tetrahedron method by values in the 34 meV to 6 meV interval when varying $\eta$ from 50 meV to 1 meV, respectively. Increasing the $\mb{q}-$mesh density to $N=192$, the interval is even smaller, going from 6 meV to 1 meV when varying $\eta$ from 50 meV to 1 meV. 
The convergence of the imaginary part at the CBM is more problematic
(see Fig.~\ref{fig:ImSigmaConvCBM}):
the value of $\Im m \Sigma_{\text{CBM}}(\varepsilon^0_{\text{CBM}})$ at $T=300$ K with the tetrahedron method is very small, around $10^{-4}$ eV, i.e. smaller than the values of $\eta$. Further convergence would require $\eta \leq 10^{-4}$ and even denser grids, which are not practical or indeed necessary.
Selecting $\eta=5$ meV and $N=128$ (values that will be used later on), the $\Im m \Sigma_{\text{CBM}} (\varepsilon^0_{\text{CBM}})$ is within an order of magnitude of the very small tetrahedron value.
Lowering $\eta$ systematically
decreases $\Im m \Sigma_{\text{CBM}}(\varepsilon^0_{\text{CBM}})$, but it also increases numerical noise as detailed below. 

\begin{figure}[htb]
    \centering
    \includegraphics[width=\columnwidth]{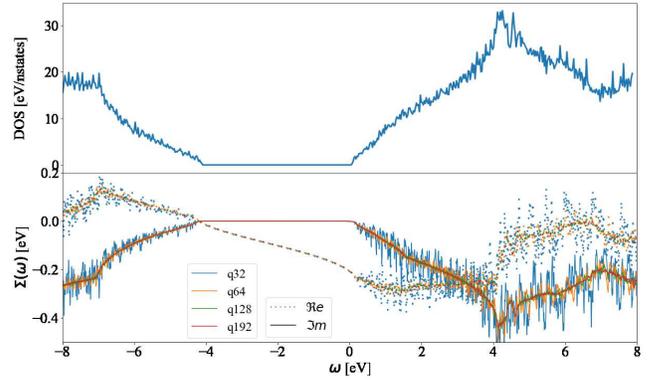}
    \caption{The a) density of states and b) self-energy at CBM and 300 K calculated using DM and $\eta = $ 5 meV for $N= 32, 64, 128$ and $192$ at CBM. The DOS is plotted as a reference, showing its similarity with the imaginary part of the self-energy. Self-energy becomes smooth and converged at $N= 128$ in green (difficult to visualize) with very small oscillations around the $N=192$ data in red (easy to visualize).}
    \label{fig:Sigma_Convwmesh_DOS}
\end{figure}

The effect of increasing the density of the $\mb{q}$-mesh in the imaginary part of the  dynamical self-energy can be observed in Fig. \ref{fig:Sigma_Convwmesh_DOS}: as the $\mb{q}-$mesh increases, the noise decreases.
We choose a $\mb{q}$-mesh grid of $N=128$ for all the following calculations, which produces low computational noise and a convergence error of 0.6\% on the value of the self energy at the CBM. 

In Fig.~\ref{fig:Sigma_Convwmesh_DOS}, we also display the density of states (DOS). The imaginary part of the self-energy can be seen as a similar sum over accessible electronic states at $m, \vec{k+q}$, but weighted by finite e-ph matrix elements, Bose-Einstein factors at finite $T$, and shifted by phonon frequencies. 
Another observation is a linear departure of $\Im m \Sigma_{\text{CBM}}(\omega)$ at $\omega_{\text{LO}}$ from the value $\Im m \Sigma_{\text{CBM}}(\varepsilon^0_{\text{CBM}})$.
This is at variance with the case of Fr\"ohlich model and polar materials\cite{Nery2018}, in which the e-ph matrix elements diverge at small $\mathbf{q}$. This results in a peak in the self-energy at the LO phonon frequency, which is not present in the DOS.
The real part of the self-energy is connected to the imaginary part by causality through the Kramers-Kronig relations.

\begin{figure}[H]
    \centering
    \includegraphics[width=\columnwidth]{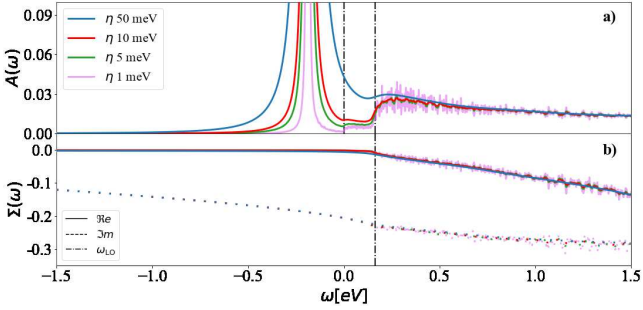}
    \caption{The a) spectral function and b) self-energy at the CBM and 300 K calculated using DM and a $N=128$ for different values of $\eta= 50, 10, 5$ and $1$ meV. The first vertical line is at 0 energy (CBM) and the second at the highest phonon frequency. The finite temperature allows for absorption of acoustic phonons, which produces a small plateau between the QP peak and the main (optical) satellite feature.}
    \label{fig:Aw_Sigma_eta}
\end{figure}

For a fixed, accurate, $\mb{q}-$mesh size ($N=128$), we show the effect of $\eta$ in Fig. \ref{fig:Aw_Sigma_eta}. At very low $\eta=$ 1 meV, there are strong oscillations in the spectral function satellite features. The width of the QP peak also increases with $\eta$, such that a numerical compromise must be found.  
Above the CBM QP energy of -0.180 eV, there are two plateaus in the spectral function: a small one starting at zero frequency, and a more important one starting at the highest phonon frequency, $\omega_{LO}$, (dot-dashed line), analogous to the peak observed in the Fr\"{o}lich self energy form for polar materials. 
To avoid computational noise, and too dense $\mb{q}-$meshes, we select for the rest of the paper a $N=128$ $\mb{q}-$mesh and $\eta=$ 5 meV. 
Analysis of the plateaus and the spectral function will be detailed in Section \ref{subsec:SpectralFunctionARPESTemperature}.

Other convergence to consider in the self-energy is the sum over $m$ bands in eq. (\ref{eq:Fan}), which includes unoccupied states and its numerical convergence over empty bands becomes burdensome. An alternative solution is replacing high-energy bands with the solution of a non-self-consistent Sternheimer equation\citep{Gonze2010}. Study of this convergence can be found in ESI section \ref{sec:Sternheimer}.
The calculation of the cumulant function at low temperatures is complicated by the very small value of $\Im m \Sigma_{\text{CBM}}(\varepsilon^0_{\text{CBM}})$. $\Im m \Sigma_{\text{CBM}}(\varepsilon^0_{\text{CBM}})$ is related to the decay of the QP and its lifetime: the smaller it is, the longer it takes the QP to decay. The range of the time mesh can be estimated through Eq. (\ref{eq:GreenFunctionInftyLimit}), which is an envelope of the Green's function, which can be used to determine the time at which QP decays and reaches a given tolerance tol, $t_{\text{max}} = - \left(\ln (\text{tol} ) + w\right) /  \Gamma$.

At low temperatures, e.g. 40 K, $\Im m \Sigma_{\text{CBM}}(\varepsilon^0_{\text{CBM}})$ is $-4.95 \times 10^{-6}$ eV and the number of frequency points required is huge. The grid size can be estimated as $\Delta \omega/\delta \omega$, where $\delta \omega$ is the spacing between points and $\Delta \omega$ is the frequency range. With $t_{\text{max}}\approx 918$ ps for a tol of $10^{-3}$ we obtain $\delta \omega = 2 \pi / t_{max}$. For a range $\Delta \omega$ of 20 eV (The convergence of $\Delta \omega$ is explained in ESI \ref{subsec:Kramers-Kronig}) we obtain 5.3M points. Therefore, we limit the calculations to room temperature and above, where considerably fewer frequency points are required. 

At $T=300$ K, $\Im m \Sigma_{\text{CBM}}(\varepsilon^0_{\text{CBM}})=-1.51 \times 10^{-4}$ eV gives a $t_{max}\approx20$ ps with a tolerance of $10^{-3}$, leading to 174.7k frequency points. Increasing the temperature, $\Im m \Sigma_{\text{CBM}}(\varepsilon^0_{\text{CBM}})$ increases, which means shorter QP lifetimes and a lower number of frequency points.
\begin{figure}[h]
    \includegraphics[width=\columnwidth]{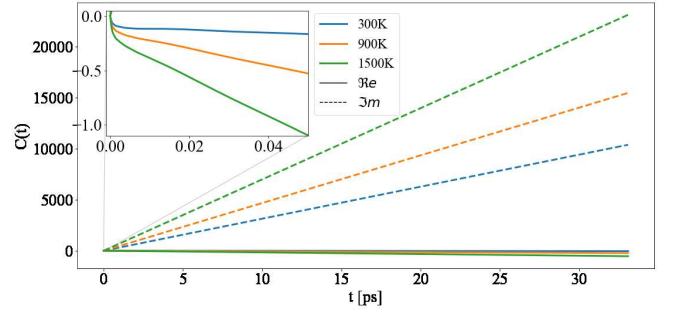}
    \caption{The cumulant function for CBM between 0 and 32 ps, for different temperatures: 300, 900, and 1500 K. The solid and the dashed lines are the real and imaginary parts of the cumulant function, respectively. The inset shows the non-linear behaviour of $\Re e C(t)$ at small $t$. } 
    \label{fig:Ct_difftmaxSlope}
\end{figure}
In Fig. \ref{fig:Ct_difftmaxSlope}, the cumulant function is shown for $T=300, 900$, and $1500$ K.
The linear behavior of the cumulant at infinite time is present both in the real and imaginary parts. The large time slope of the real part is $\Im m \Sigma_{\text{CBM}}(\varepsilon^0_{\text{CBM}})$ and the slope of the imaginary part is $\Re e \Sigma_{\text{CBM}}(\varepsilon^0_{\text{CBM}})$. The intercept of $\Re e C(t)$, $w_k$, originates from the asymmetry\citep{Gumhalter2005} of the denominator of $\Sigma$ when summing over all $j \mb{q}$ states, both elements described in Eq. (\ref{eq:GreenFunctionInftyLimit}) and its convergence is examined in more detail in the ESI \ref{subsec:Kramers-Kronig}. The increase of the $\Re e C(t)$ at $t\to \infty$ accelerates the decay of the QP and determines its lifetime $\tau = 1/2 \Gamma$. 

\hfill
\subsection{Spectral function and ARPES at finite temperature}\label{subsec:SpectralFunctionARPESTemperature}

\begin{figure}[h]
    \centering
    \includegraphics[width=\columnwidth]{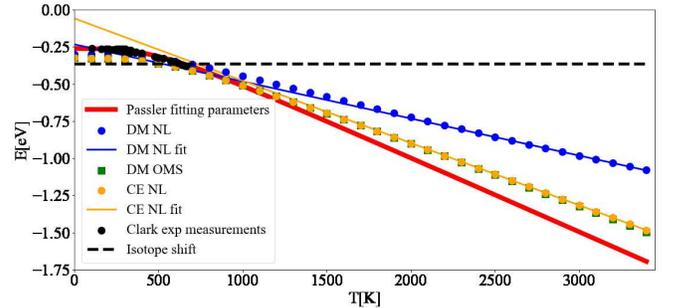}
    \caption{Variation of the indirect band gap of diamond with respect to temperature. The bare band gap was set to zero. The calculations of the band gap using DM-NL (blue points) do not produce a strictly linear behaviour, and the DM-OMS (green squares) produces the same values as the CE using the NL approach (yellow points). P\"assler fit \cite{Passler1999} (solid red line) to the experimental points by Clark \textit{et al} \cite{Clark1964} (black points) and measured ZPR by isotope shift\citep{Cardona2005} (black dashed line). The calculations were done with $N=64$ and $\eta = 10$ meV.} 
    \label{fig:ReSigma_T2}
\end{figure}

After converging the self-energy, we can evaluate the spectral function at finite temperatures using the DM and the CE approaches. 
In Fig. \ref{fig:ReSigma_T2}, the ZPR can be observed by following the shift of the renormalized band gap at $T=0K$ relative to the bare band gap, set to zero. The ZPR can be determined experimentally by extrapolating the linear regime at high temperatures to 0 K. Using a P\"assler fit\citep{Passler1999} to the measurements of Clark \textit{et al}\citep{Clark1964} gives a ZPR of -0.259 eV. Using also the difference in renormalization for isotopes $^{12}$C and $^{13}$C, the obtained ZPR is -0.364 eV\citep{Cardona2005}. Table \ref{tab:zpr} shows a range of calculated ZPR that go from -315 meV to -619 meV.

\begin{table*}
\begin{tabular*}{\textwidth}{@{\extracolsep{\fill}}cccccccc}
ZPR (meV) & Calc/Exp & DF(P)T/MC & SC/AHC & (non-)A & EE & TE &(An)Ha \\ \hline
-364\citep{Cardona2005} & Exp & & & & & \\
-325\hspace{8pt}   & FP(This work) & DFPT & AHC & non-A & No & No & Ha \\
\hline
-315\citep{Karsai2018} & FP & MC & SC & A  & No & Yes & ? \\
-320\citep{Antonius2015} & FP & DFT & SC & A & No & No & AnHa \\
-330\citep{Ponce2015} & FP & DFPT & AHC & non-A & No & No & Ha \\
-337\citep{Karsai2018} & FP & MC & SC & A & G$_0$W$_0$ & Yes & ? \\
-366\citep{Antonius2015} & FP & DFPT & AHC & non-A & No & No & Ha \\
-372\citep{Antonius2015} & FP & DFPT & AHC & A & No & No & Ha \\
-380\citep{Ponce2015} & FP & DFPT & AHC & A & No & No & Ha \\
-436\citep{Antonius2015} & FP & DFPT & SC & A & No & No & Ha \\
-439\citep{Antonius2015} & FP & DFT & SC & A & No & No & Ha \\
-462\citep{Monserrat2014} & FP & DFT & SC & A & No & No & Ha \\
-619\citep{Zollner1992}& Semi-Emp$^{*}$  & DFPT & AHC & A & No & Yes & Ha \\
 \end{tabular*}
 \caption{Diamond indirect band-gap ZPR calculations found in the literature. Meaning of the abbreviations: All calculations were done within first-principles (FP), except the last one $^{*}$ which used a semi-empirical exchange-correlation functional (Semi-Emp.); Allen, Heine and Cardona approach (AHC); Supercell (SC) calculations; Thermal expansion of the crystal (TE); Calculations with electron-electron (EE) interactions; Frozen-phonons approximation (FP); Harmonic approximation (HA); Anharmonic effect (AnHA); (?) Not explicitly clarified in the references, but presumably harmonic.
} 
 \label{tab:zpr}
\end{table*}

The fast convergence of the real part of the self-energy allows to determine the renormalization of the band gap in Fig. \ref{fig:ReSigma_T2} at a coarse $N = 64$ $\mb{q}-$mesh and at a $\eta$ of 10 meV.
The slope of the DM-OMS energy is equivalent to the CE-NL energy with -0.409 meVK$^{-1}$, giving an extrapolated ZPR from high temperatures of -0.281 eV. 
This value differs from the CE-NL calculated at T=0 K by -0.066 eV. This discrepancy emerges from two factors: the linear extrapolation should be made above diamond's high Debye temperature (2246 K)\citep{Burk1958}, and the non-adiabatic self-energy, Eq. \ref{eq:Fan}, includes a plus and minus term of the phonon energy that is not present in the adiabatic self-energy (an adiabatic self-energy yields the same ZPR when calculated at $T = 0$ K and when extrapolated from high temperatures).
The very similar behaviour with temperature for DM-OMS and CE-NL derives from the cancellation of errors of the former between the exclusion of higher order e-ph interactions and the evaluation of the self-energy at the KS energy. The behavior of DM-NL is erratic, since it is not linear at high temperatures, as opposed to CE-NL and DM-OMS. If one attemps a linear extrapolation for temperatures above 2500 K, where it becomes almost linear, the ZPR gives -0.092 eV
with a discrepancy of -0.207 eV from the ZPR calculated at T=0K, three fold the CE discrepancy.

\begin{figure}[h]
    \centering
    \includegraphics[width=\columnwidth]{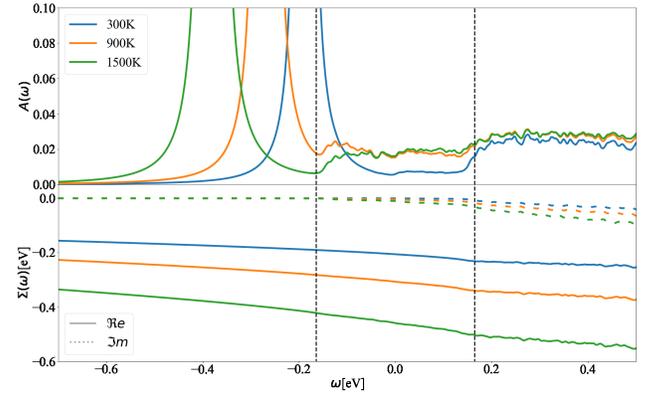}
        \caption{(a) The DM spectral function and (b) the real and imaginary parts of the self-energy at the CBM for T= 300, 900 and 1500 K, with $N=128$ and $\eta = 5$ meV. The vertical dashed black lines correspond to the phonon energies $-\omega_{LO}$ and $\omega_{LO}$ around the KS energy that is used as origin of the x-axis. }
    \label{fig:SelfEnergy_Temperature}
\end{figure}

Going further by evaluating the dynamical part of the spectral function at the CBM in Fig. \ref{fig:SelfEnergy_Temperature} for T=300, 900, and 1500 K, one can observe a temperature-dependent offset in the QP peak (the KS energy is $\omega = 0$) which corresponds to the DM-NL energy.
Unlike the Fr\"{o}hlich polaron in simple polar materials, where there are satellite peaks with well-defined replica, here the signature of a nonpolaron is a plateau. As the temperature increases, more states become available; in particular, $\Sm$ starts to contribute to the presence of another plateau with states below the KS energy.
Similar to polar materials, the DM approach overestimates the energy distance between the main QP and the high plateau, giving 2.4 $\omega_{LO}$ at T=300K. % instead of $\omega_{LO}$.

\begin{figure}[h]
    \centering
    \includegraphics[width=\columnwidth]{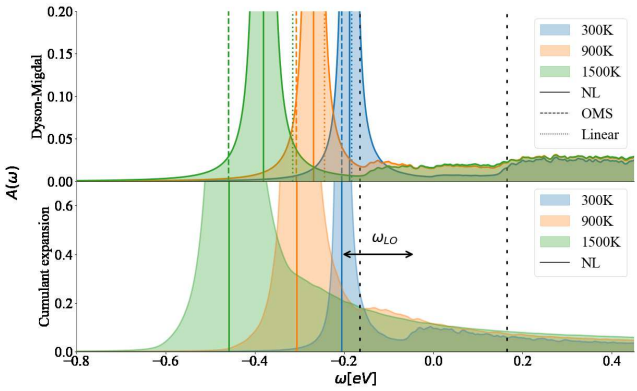}
    \caption{Spectral function calculated at the CBM at 300, 900, 1500 K using DM and CE methods. The vertical colored lines show the QP energy calculated using non-linear (NL), on-the-mass-shell (OMS) and linear approaches. The vertical dashed black lines correspond to $\pm\omega_{LO}$. $\eta$ is set to 5 meV and $N=128$.}
    \label{fig:AwDMandCumulant_Diamond}
\end{figure}

\begin{figure}[h]
    \centering
    \includegraphics[width=\columnwidth]{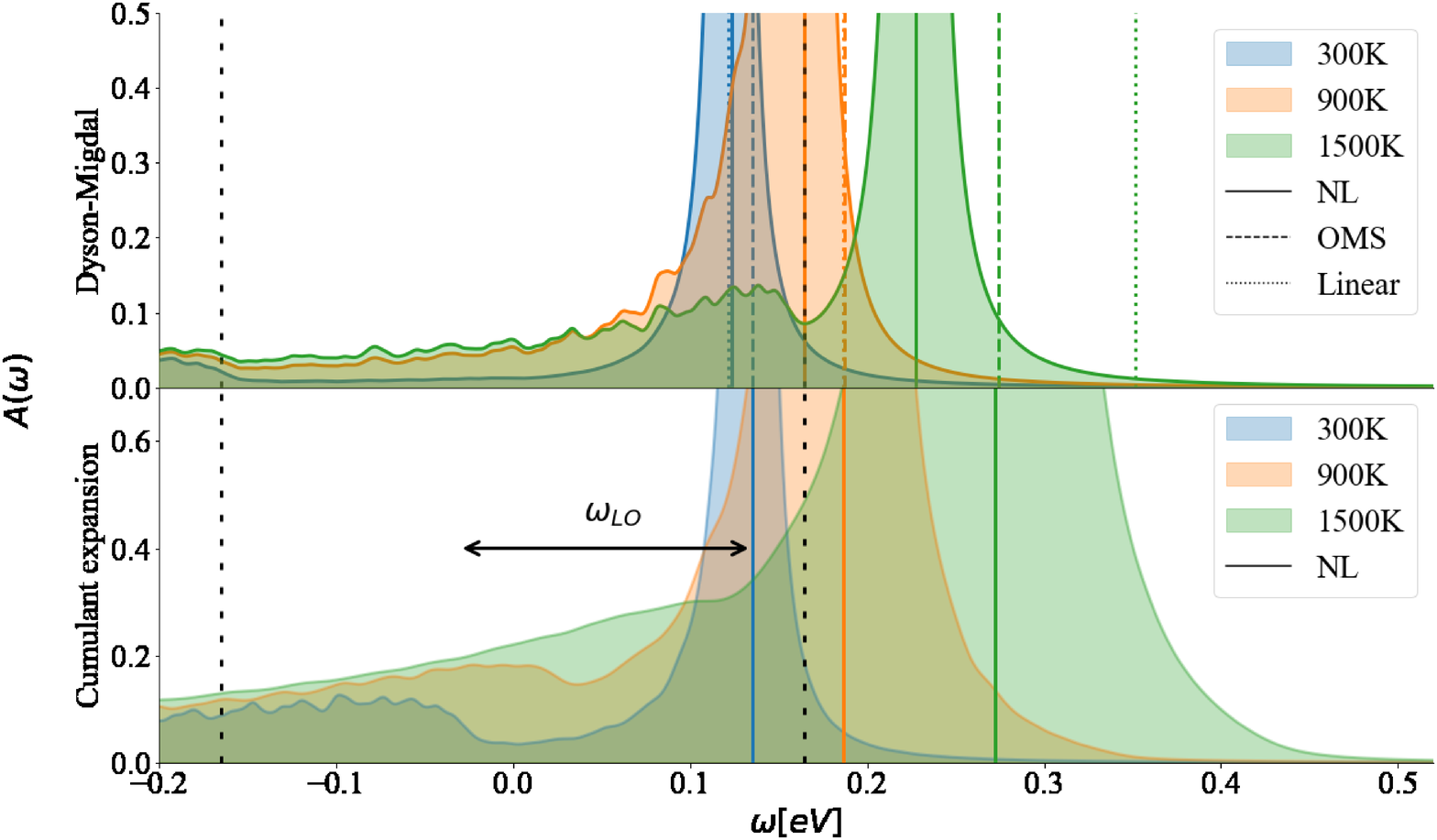}
    \caption{Spectral function calculated at the VBM at 300, 900, 1500 (K) using DM and CE methods. The vertical colored lines show the QP energy calculated using non-linear (NL), on-the-mass-shell (OMS) and linear approaches. The vertical dashed black lines correspond to $\pm\omega_{LO}$. $\eta$ is set to 5 meV and $N=128$.}
    \label{fig:Aw_DM_CE_VBM}
\end{figure}

The energy shift of the QP relative to the KS energy at both the CBM and VBM reduces the band gap, as shown in Figs.~\ref{fig:AwDMandCumulant_Diamond} and  \ref{fig:Aw_DM_CE_VBM}, respectively (the KS energy is set at 0 eV in each case). Taken together, they determine the band gap shift in Fig.~\ref{fig:ReSigma_T2}. The QP peaks are located at the $\varepsilon^{NL}$ energy. The DM-NL energy differs from the CE-NL more noticeably at higher temperatures. Since the latter contains higher orders of e-ph interactions, the CE approach is more appropriate to determine the QP energy. These energies are very close to the DM+OMS approximation, as also observed in Fig.~\ref{fig:ReSigma_T2}.

At the DM level, the frequencies of both plateaus are disconnected from the QP energy, as discussed in the ESI \ref{subsubsec:SigmaPlusSigmaMinus}. As the temperature increases, the nonpolaron plateau seems to get closer to the QP peak. However, this is an artifact of DM, as new states become accessible by $\Sm$. In DM, $\Sp$ gives contributions that start at $+\omega_{LO}$ counting from the \emph{KS energy}, independently of the temperature or position of the QP peak. At higher temperatures, $\Sm$ becomes larger and contributes to the spectral function at $-\omega_{LO}$. Since the band renormalization is larger than $-\omega_{LO}$, the $-\omega_{LO}$ plateau still appears to the right of the QP peak, giving the appearance that the plateau is shifting with temperature, while it actually corresponds to another plateau. There is an un-intuitive behaviour with DM at the VBM, where the broadening of the QP at 900 K seems wider than at 1500 K. This is due to the overlap between the states created by $\Sp$ at $+\omega_{LO}$ (counting from $\omega=0$) and the QP peak. More detailed analysis is deferred to ESI \ref{subsubsec:SigmaPlusSigmaMinus}.

For the CE approach, let us focus first on the CBM. At low temperatures, $\Sp$ produces a plateau at the right energy distance, $+\omega_{LO}$, from the QP peak. When increasing the temperature, the main QP becomes wider. 
It is easier to identify features in the DM spectral function, because the self-energy consists only of two terms, $\Sp$ and $\Sm$. In the CE instead, as it consists of an exponential representation, higher order terms mix the contributions from the QP peak with satellite features in the spectral function. This smooths out the plateau, which is no longer separate from the QP at higher temperatures, although a weak peak is visible at 900 K, and a long tail is still present at 900 K and 1500 K. Results are similar for the VBM (Fig.~\ref{fig:Aw_DM_CE_VBM}).
Unlike polar insulators, such as LiF and MgO\citep{Nery2018}, or the Fr\" ohlich model in the CE approach, there is no visible signature at $+2\omega_{LO}$.

\begin{figure}[!htpb]
    \centering
    \includegraphics[width=\columnwidth]{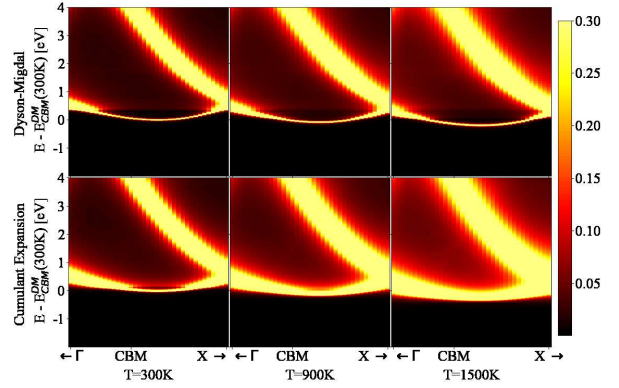}
    \caption{Calculated ARPES spectra using DM (top) and CE (bottom) at the bottom of the conduction band. Zero was set to the CBM energy in the DM approach. The calculations were done at temperatures equal to 300, 900 and 1500 K. To be able to observe the presence of the additional signature present in the spectral function at VBM and CBM (a line just above the main QP peak on the CBM), the intensity scale of the density of states (in colors) was limited to 0.3. The plateau is visible for the different temperatures in the DM approach and the broadening changes slightly. This is opposite to CE, where we can only see the plateau at 300K, due to mixing between the broader QP and the nonpolaronic feature. }
    \label{fig:ARPES_CBM}
\end{figure}

The spectral function close to the band edges was calculated along a path between high-symmetry points in reciprocal space, in order to reproduce an ARPES experiment. This allows us to visualize the effects of phonons, and to compare the DM and CE approaches in reciprocal space (see Fig \ref{fig:ARPES_CBM}).
For both approaches we can perceive at 300 K a light red color just above the yellow band at the CBM, which represents the nonpolaronic signature. In DM, the plateau  is located at 2.4 $\omega_{\text{LO}}$ above the CBM quasi-particle peak, and as the temperature increases, the plateau shifts down about 0.32 eV, which seems constant in this energy interval (the same shift can be seen in the top plot of Fig. \ref{fig:AwDMandCumulant_Diamond} with the shift of the plateau to lower energy from 300 K to 900 K).
The QP peak shifts down from -0.179 eV to -0.367 eV when the temperature increases from 300 K to 1500 K, respectively, see Table \ref{tab:ValuesEnergyTemperature}. 
In the CE, the satellite band is located $\omega_{\text{LO}}$ away from the CBM QP peak at 300 K. As temperature increases the QP mixes with the nonpolaronic signature and increases the broadening. There is also a huge broadening increase at the degenerate bands close to the high-symmetry point X. The changes with temperature and absolute energy resolution mean these features should be visible experimentally, and we hope to stimulate more detailed ARPES studies on intrinsic diamond as a model for nonpolarons.

\begin{figure}[!htpb]
     \centering
     \begin{subfigure}[b]{\columnwidth}
         \centering
         \includegraphics[width=\textwidth]{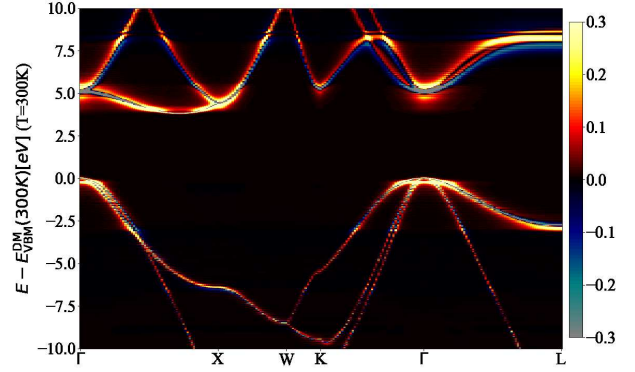}
         \caption{$T=300K$}
         \label{fig:ARPES300K_diff}
     \end{subfigure}
     \begin{subfigure}[b]{\columnwidth}
         \centering
         \includegraphics[width=\textwidth]{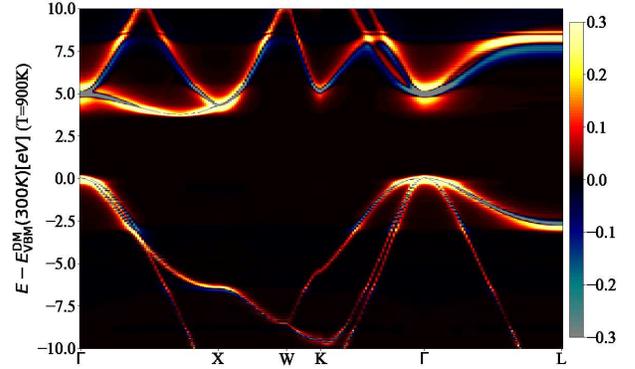}
         \caption{$T=900K$}
         \label{fig:ARPES900K_diff}
     \end{subfigure}
     \begin{subfigure}[b]{\columnwidth}
         \centering
         \includegraphics[width=\textwidth]{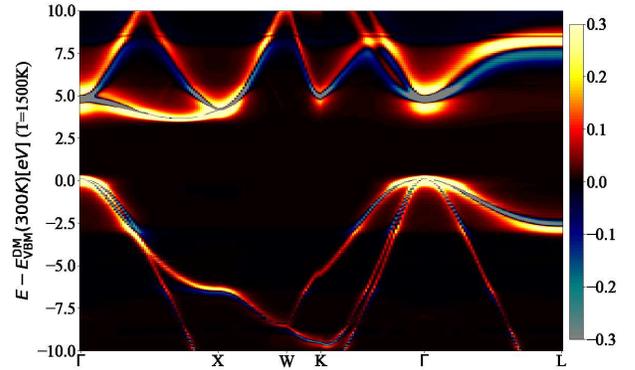}
         \caption{$T=1500K$}
         \label{fig:ARPES1500K_diff}
     \end{subfigure}
        \caption{Calculated spectral function difference between the CE and DM, $A^\mathrm{CE}(\o) - A^\mathrm{DM}(\o)$  at a) 300 K, b) 900 K, and c) 1500 K. To highlight the nonpolaronic signature near the VBM and CBM, the intensity scale of the density of states (colormap) was limited to $\pm 0.3$. For positive values (red and yellow colors), the CE has more intensity than DM, and the opposite happens for the negative values (blue and grey colors). Comparatively with the narrow bands of DM, the CE approach has a much broader spread of the QP weight. 
        Around most of the high symmetry k-points the broadening is almost symmetric, except L, where it is asymmetric, creating a tail. 
         }
        \label{fig:ARPES_diff}
\end{figure}

To compare more quantitatively, the DM calculated ARPES was subtracted from the CE calculations in Fig \ref{fig:ARPES_diff}. The intensity range is limited between $-0.3$ and $0.3$ to detail the view of the nonpolaronic signatures. 
For positive values the CE has a higher spectral function, while the opposite occurs for negative values. One can observe a broadening or/and shift effect when transiting from DM (a thin line green or grey) to CE, which surrounds the green or grey line by a yellow or red area. 
In some parts of the band structure, such as the bottom of the conduction band close to the high symmetry point $L$, one can observe a tail to higher energies and an asymmetry close to the QP peak, as the weight at the QP energy using DM (blue), is spread to the tail within CE (yellow). As the temperature increases from 300 K in Fig. \ref{fig:ARPES300K_diff}, 900 K in Fig. \ref{fig:ARPES900K_diff}, to 1500 K in Fig. \ref{fig:ARPES1500K_diff}, the broadening increases, especially for states close to the Fermi level. The shift of the QP peak from DM to CE as seen in Figs. \ref{fig:AwDMandCumulant_Diamond} and \ref{fig:Aw_DM_CE_VBM} is not visible in the full band structure, as the scale of the energy in the ARPES is much wider than the energy shift.

\subsection{Contribution due to long-range and short-range fields}\label{subsec:Quadrupolemodel}

\begin{figure}[h!]
    \centering
    \includegraphics[width=\columnwidth]{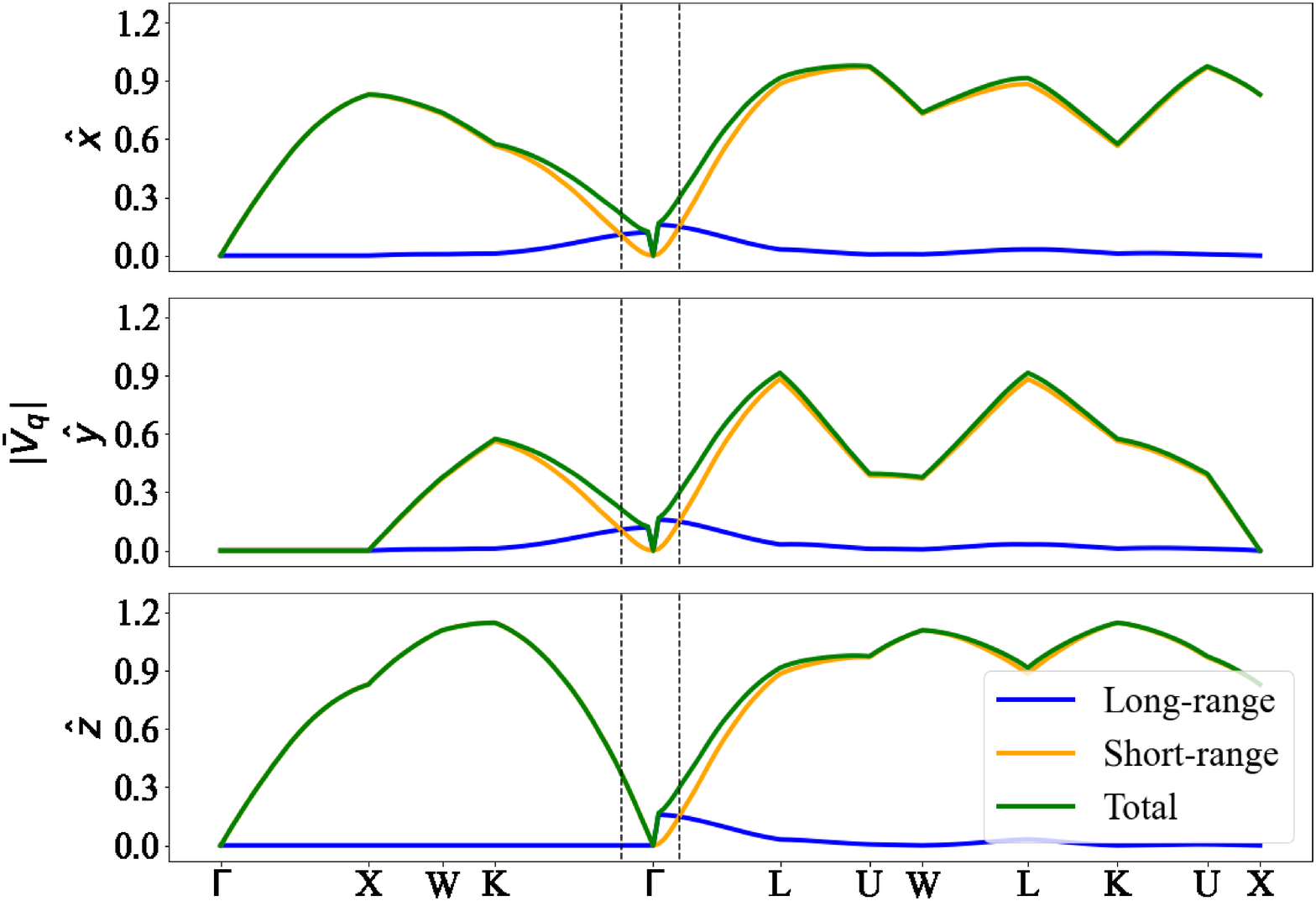}
    \caption{Unit-cell average of the total DFPT scattering potential (green) compared with the long-range part (blue) and the short-range contribution (orange). 
    The three subplots corresponds to the e-ph potential generated by the displacement of a C atom
    along one of the three reduced directions $\hat{x}$, $\hat{y}$, $\hat{z}$. Both atoms present the same $\left| \bar{V}_{\mb{q}}\right|$.
    }
    \label{fig:Potential_Quad_all}
\end{figure}

In this section, we employ the Fourier-based interpolation technique discussed in Section~\ref{subsec:QuadrupolesMethods} to analyze the
contribution given by the SR and LR quadrupolar fields to the e-ph scattering potentials.
More specifically, we compare 
the unit cell average of the (local part) of the e-ph scattering potential~\cite{Brunin2020_2}
\begin{equation}
    \bar{V}_{\kappa\alpha,\qb} =
    \dfrac{1}{\Omega}\int_\Omega \dd\rr\,V_{\kappa\alpha,\qb}(\rb) e^{-i\qq\cdot\rr},
\end{equation}
for several $\mb{q}$-points along a high-symmetry path. 
Three different approaches are used to dissect the contributions to $\bar{V}$.
In the first one, the LR contribution is obtained by 
approximating $V_{\kappa\alpha,\qb}(\rr)$ with 
the model in Eq.\ref{eq:VooglPotential}, and dynamical quadrupoles are given by
\begin{equation}
    Q_{\kappa\alpha}^{\beta\gamma} = (-1)^{\kappa+1} Q_\kappa |\varepsilon_{\beta\gamma\alpha}|,
\end{equation}
with $\epsilon$ the Levi-Civita tensor, $\kappa$ the site index and $Q_\kappa = 2.52$. 
In the second approach, the SR part is obtained by Fourier-interpolating\citep{Brunin2020_2}
the short-ranged $W_{\kappa,\alpha}(\rr, \RR)$, 
in which quadrupole fields have been removed.
Finally, the total e-ph scattering potential is obtained by performing explicit DFPT calculations for all the $\qq$-points of the path.
Figure~\ref{fig:Potential_Quad_all} compares the results obtained with these three approaches.
Note, in particular, the jump discontinuity in the DFPT values 
for $\qq \rightarrow \Gamma$, that is due to the LR quadrupolar potential.
As discussed in Ref.~\citen{Brunin2020}, a proper treatment of the LR quadrupole terms is necessary for an accurate interpolation of e-ph matrix elements in the long-wavelength limit.
According to our results for diamond, however, an e-ph self-energy evaluated with e-ph matrix elements obtained from the LR model potential alone (Eq. \ref{eq:VooglPotential}) cannot reproduce the results obtained with e-ph matrix elements including both the SR and the LR part.
In diamond, 
short-range crystal fields provide the most significant contribution to the QP formation
with only a small influence of long-range quadrupoles.
We can estimate the real space distance beyond which quadrupolar long-range fields dominante short-range ones, by determining the point in the reciprocal lattice around $\Gamma$ where the short-range potential becomes equal to the long-range one. By doing so, one can estimate the radius of interaction between the short-range fields and the electronic carrier.
This distance is of $ \left| \mb{q} \right| \approx 0.18$ in fractional units, represented by the two vertical dashed lines in Fig \ref{fig:Potential_Quad_all}. In the direct lattice it corresponds to around 18.3 \AA, or around 7 unit cells in real-space, showing that diamond could be described as a medium polaron or larger. Note that we are determining an approximation of an interaction radius and not the size of the polaron itself. Further calculations will be carried out to estimate a polaron radius as in Refs.~\citen{Peeters1985,Sio2019}. 

Even if a quadrupole long-range field is present, it can be noticed from Fig. \ref{fig:el-phmatrixelements} that its importance disappears when calculating the scattering matrix elements, and there is thus negligible contribution from long-range fields to the nonpolaron spectral function. The maximum value of $\left|g_{nm \mb{k}}^{j \mb{q}} \right|$ is of 0.97, close to the effective coupling constant found experimentally in surface hydrogen-terminated on diamond of $1.1$ \citep{Rameau2011}.

\section{Conclusion}

First-principles e-ph calculations for diamond reveal spectroscopic signatures that originate from a quasiparticle that can be called nonpolaron: a bound electron state dressed with phonons in a non-polar system. We expect such signatures to be generic and occur in other covalent crystals. 

Unexpectedly, in diamond the contribution to the formation of the nonpolaron from the long-range quadrupole fields is negligible and the binding is mainly generated by the short-range fields.

The calculations were done using both the standard Dyson-Migdal approach and the cumulant expansion. According to the CE, the signature of a nonpolaron is a plateau in the spectral function, starting one phonon energy above the main QP peak (the phonon is the largest frequency one). 
The plateau shape is at variance with the satellite peaks seen in polar systems, each separated from the QP peak or the previous replica by the LO phonon energy. 

As in polar materials, we find that the QP is not properly captured in one shot NL-DM, 
though it might be improved by self-consistency in G. 
We show that this results yields an incorrect temperature dependence of the DM QP peak which is not linear at high temperatures. The DM approximation also yields a wrong position for the plateau structure, which is located at $\pm \o_\mathrm{LO}$ relative to the KS value, independently of the position of the QP. 

Calculations of the ZPR within CE and using the DM on-the-mass-shell approximation give similar results, and are in good agreement with experimental values for the ZPR. 
In addition,
the cumulant expansion method delivers a physically correct nonpolaron energy, and the inclusion of the higher order terms in the e-ph interaction provides a more "realistic" view of the spectral function. The calculated ARPES spectrum of diamond around the band gap shows a broader dispersion within CE compared to Dyson-Migdal.

Finally, in order to obtain the spectral functions, self-energies, ZPR and temperature dependence of the band gap, convergence with respect to the $\mb{q}$-mesh density points and the broadening parameter $\eta$ were carefully studied.
The real part of the QP self-energy converges quickly in opposition to the imaginary part. The narrow QP broadening at the VBM and CBM implies a very strong sensitivity of the imaginary part of the self-energy to convergence, and the imaginary $\eta$ introduced in the Green's function should be chosen carefully. Our study showcases the combination of convergence techniques to reduce computational cost, in particular: the Sternheimer equation which substitutes empty bands by a linear response equation, and the Kramers-Kronig relation, which replaces a difficult reciprocal space sum by the use of a pre-calculated complex self-energy.

This paper motivates further research, both experimental and theoretical, on the topic of non-polar phonon modes in interacting electron-phonon systems.

\FloatBarrier
\section*{Author Contributions}
J.C.A. prepared the manuscript, carried out calculations,
prepared all graphical material and implemented the cumulant expansion, with help from M.G., J.P.N. and M.J.V.. J.P.N. contributed by including supplementary information, discussions, editing and reviewing. M.G. and X.G. contributed with discussions, editing and reviewing.  M.J.V. and X.G. acquired the main funding for the project. M.J.V. supervised the project and calculations, and edited and reviewed the manuscript.

\section*{Conflicts of interest}
The authors have no conflicts of interest to declare.

\section*{Acknowledgements}

This work has been supported by the Fonds de la Recherche Scientifique (FRS-FNRS Belgium) through
the PdR Grant No. T.0103.19 - ALPS.
Computational resources have been provided by 
CECI funded by the FRS-FNRS Belgium under Grant No. 2.5020.11,
as well as the Tier-1 supercomputer of the F\'ed\'eration Wallonie-Bruxelles, 
infrastructure funded by the Walloon Region under grant agreement No. 1117545. 
We acknowledge a PRACE award granting access to MareNostrum4 at Barcelona Supercomputing Center (BSC), Spain (OptoSpin project id. 2020225411).

\bibliographystyle{ieeetr}
\bibliography{literature}

\balance

\onecolumn

  \begin{@twocolumnfalse}
%{\includegraphics[height=30pt]{head_foot/PCCP}\hfill\raisebox{0pt}[0pt][0pt]{\includegraphics[height=55pt]{head_foot/RSC_LOGO_CMYK}}\\[1ex]
%\includegraphics[width=18.5cm]{head_foot/header_bar}}\par
\vspace{1em}
%\sffamily
%\begin{tabular}{m{4.5cm} p{13.5cm} }
{\centering
\noindent\LARGE{\textbf{
Supplementary Information of
%Nonpolarons: Polarons in non-polar systems
Spectroscopic signatures of nonpolarons : the case of diamond}} % \\
\vspace{0.3cm}% & \vspace{0.3cm} \\

 \noindent\large{Joao C. de Abreu,\textit{$^{1}$} Jean Paul Nery,\textit{$^{2}$} Matteo Giantomassi,\textit{$^{3}$} Xavier Gonze\textit{$^{3,4}$} and Matthieu J. Verstraete\textit{$^{1}$}}% \\

\vspace{1 em}

\textit{$^{1}$~nanomat/Q-MAT/CESAM and European Theoretical Spectroscopy Facility, Universit\'e de Li\`ege, B-4000 Belgium}

\textit{$^{2}$~Dipartimento di Fisica, Universit\`a di Roma La Sapienza, I-00185 Roma, Italy }

\textit{$^{3}$~UCLouvain, Institute of Condensed Matter and Nanosciences (IMCN), Chemin des \'Etoiles~8, B-1348 Louvain-la-Neuve, Belgium }

\textit{$^{4}$ Skolkovo Institute of Science and Technology, Moscow, Russia }

%\textit{ $^{4}$~Skolkovo Institute of Science and Technology, Moscow, Russia }
}
\vspace{2 em}
%\includegraphics{head_foot/dates} & 

% \vspace{1.0cm}

\renewcommand{\thesection}{S.\arabic{section}}
\numberwithin{equation}{section}
\numberwithin{figure}{section}
\numberwithin{table}{section}

\setcounter{section}{0}
\setcounter{equation}{0}
\setcounter{figure}{0}
\setcounter{table}{0}

   \section{Electronic and phonon structure} \label{subsec: Electronic Phonon}
   
   The non-interacting wave-functions $\psi_{n \mb{k} }^{0}$ and eigenvalues $\varepsilon_{n \mb{k} }^{0}$ are determined self-consistently from the Kohn-Sham (KS)\citep{KohnSham1965} equation in density functional theory (DFT)\citep{Martin2004},
%    \begin{equation} \label{eq:KSeq}
%     \begin{split}
%           [ T [ \rho (\mb{r})] + v_{\text{ext}} (\mb{r}) &  + v_H [ \rho(\mb{r})] \\
%         & \left. + v_{\text{XC}} [ \rho (\mb{r})] \right]  \psi_{n \mb{k} }^0 (\mb{r}) = \varepsilon^0_{n \mb{k} } \psi^0_{n \mb{k} } ( \mb{r})
%     \end{split}
%   \end{equation}
    \begin{equation} \label{eq:KSeq}
           [ T + v_{\text{ext}} (\mb{r})   + v_H [ \rho(\mb{r})] \left. + v_{\text{XC}} [ \rho (\mb{r})] \right]  \psi_{n \mb{k} }^0 (\mb{r}) = \varepsilon^0_{n \mb{k} } \psi^0_{n \mb{k} } ( \mb{r})
   \end{equation}
   where $\rho(\mb{r}) = 
   \sum_{n \mb{k} }  f_{n \mb{k} } |\psi^0_{n \mb{k}}(\mb{r})|^2$ is the electronic density. Inside the brackets the first term is the kinetic energy operator and the other terms constitutes the KS potential, $v^{KS}$, composed of (from left to right): the external potential, the Hartree energy potential, and the exchange-correlation potential. 
   %For practical calculations, the main approximations are the substitution of the external potential by pseudo-potentials that account for the interaction of the valence electrons with the nuclei and the core electrons, while the exchange-correlation energy is approximated here by the local density approximation (LDA)\citep{Perdew1981,Perdew1992}. 
   
   DFPT is used to calculate the interatomic forces constants, which are the second derivative of the total energy with respect to atomic displacements. The phonon frequencies $\omega_{j\mathbf{q}}$ and the eigendisplacements $U_{j\mathbf{q}}$ are thus obtained as solutions of the generalized eigenvalue problem\citep{Gonze1997} involving the interatomic forces constants $C (\mb{q})$,
   
   \begin{equation}
       \sum_{\kappa' \beta} C_{\kappa \alpha, \kappa' \beta} ( \mb{q} ) U_{j\mathbf{q}} ( \kappa' \beta) = M_\kappa \omega^2_{q} U_{j\mathbf{q}} (\kappa \alpha)
   \end{equation}
   where $M$ is the atomic mass, $\kappa$ and $\kappa'$ are the index of the ions in the unit cell, $j$ is the phonon mode, and $\alpha$ and $\beta$ are the Cartesian directions. The e-ph matrix elements are derived from the first order variation of the KS potential, and given by
   
%   \begin{equation} \label{eq:couplingconstantKSpotential} 
%   \begin{split}
%       g_{nm \mb{k}}^{j \mb{q}} = &\int d \mb{r} 
%       (\psi^0_{n \mb{k}+ \mb{q}} (\mb{r}))^+ \frac{e^{i \mb{q} \cdot \mb{r}} } { \sqrt{2 \omega_{j \mb{q}}}}  \\
%       & \times \sum_{\kappa \alpha} \frac{U_{\kappa \alpha, j}(\mb{q})} { \sqrt{M_{\kappa}}} \left( \partial_{\kappa \alpha, \mb{q}} v^{KS} (\mb{r}) \right) \psi^0_{m \mb{k}} (\mb{r})
%   \end{split}
%   \end{equation}
      \begin{equation} \label{eq:couplingconstantKSpotential} 
       g_{nm \mb{k}}^{j \mb{q}} = \int d \mb{r} 
       (\psi^0_{n \mb{k}+ \mb{q}} (\mb{r}))^+ \frac{e^{i \mb{q} \cdot \mb{r}} } { \sqrt{2 \omega_{j \mb{q}}}}  \times \sum_{\kappa \alpha} \frac{U_{\kappa \alpha, j}(\mb{q})} { \sqrt{M_{\kappa}}} \left( \partial_{\kappa \alpha, \mb{q}} v^{KS} (\mb{r}) \right) \psi^0_{m \mb{k}} (\mb{r}).
   \end{equation}
   
The first-order derivative of $v^{KS}$ is obtained by solving self-consistently a system of Sternheimer equations. All these calculations  give us the quantities necessary to compute the e-ph self-energy.\\

\section{Computational details} \label{subsecapp:Computational details}

We performed calculations for the KS band structure using the LDA approximation\citep{Perdew1981,Perdew1992}. 
Core electrons were taken into account using a norm-conserving pseudopotential\citep{vanSetten2018}. The electronic wave functions were expanded in plane-wave basis set with an energy cutoff of 40 Ha.
Phonon properties together with the e-ph scattering potentials were calculated by means of DFPT. 
and a $8\times8\times8$ grid both for electrons and phonons. 
Then we performed MBPT calculations to determine the e-ph correction to the initial LDA electronic structure. %electronic and phonon ground-states achieved a convergence of the total energy of $1$ meV with a. 
%$8\times8\times8$ Monkhorst-Pack\citep{Monkhorst1976} $\mb{k}$- and $\mb{q}$- grid sampling.
ABINIT\citep{Romero2020,Gonze2020} was used for all the calculations.

Diamond atoms have tetrahedral geometry and bond via sp$^3$ hybrid orbitals. There are two atoms per unit cell, with a relative displacement of $(1/4, 1/4, 1/4) a$, where $a$ is the lattice parameter.
The relaxed structure has $a = 3.575$ \AA, differing by $0.16$\% from the experimental value \citep{Stoupin2010}.
The calculated fundamental band gap of diamond is 4.173 eV and the direct band gap is 5.610 eV. Calculations at the LDA level are known to underestimate the band gap, in this case, the errors are around 24\% and 21\%, respectively\citep{Logothetidis1992}. 

The phonon band structure is shown in Fig. \ref{fig:Phonons_BS}. Although there is no LO-TO splitting since diamond is not polar, we refer to the largest phonon frequency at $\Gamma$ as $\omega_{LO}$ (163 meV).
%\textcolor{red}{JP: Not important, but it looks slightly larger in the plot.}.
%, to which we will name $\omega_{\text{LO}}$ as comparative to the highest phonon mode on a polar material.

\begin{figure}[H]
    \centering
    \includegraphics[width=0.8\columnwidth]{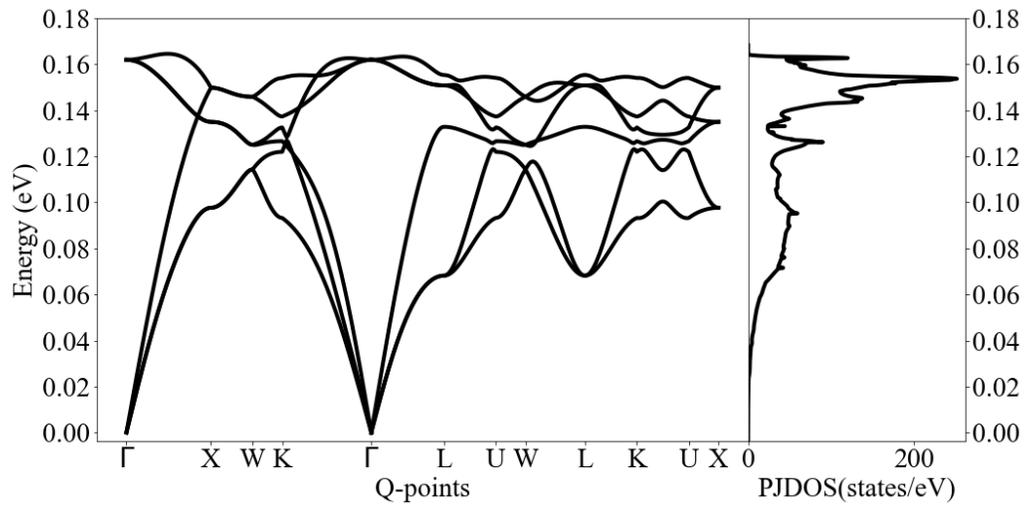}
    \caption{The phonon band structure (left) and density of states (right) of diamond obtained by Fourier-interpolating the
    dynamical matrix evaluated on a $8\times8\times8$ $\mathbf{q}$-grid.}
    % Left: Phonon band structure. Right: Phonon density of states. }
    \label{fig:Phonons_BS}
\end{figure}

\section{Kramers-Kronig}\label{subsec:Kramers-Kronig}

\begin{figure}[H]
    \centering
    \includegraphics[width=0.8\columnwidth]{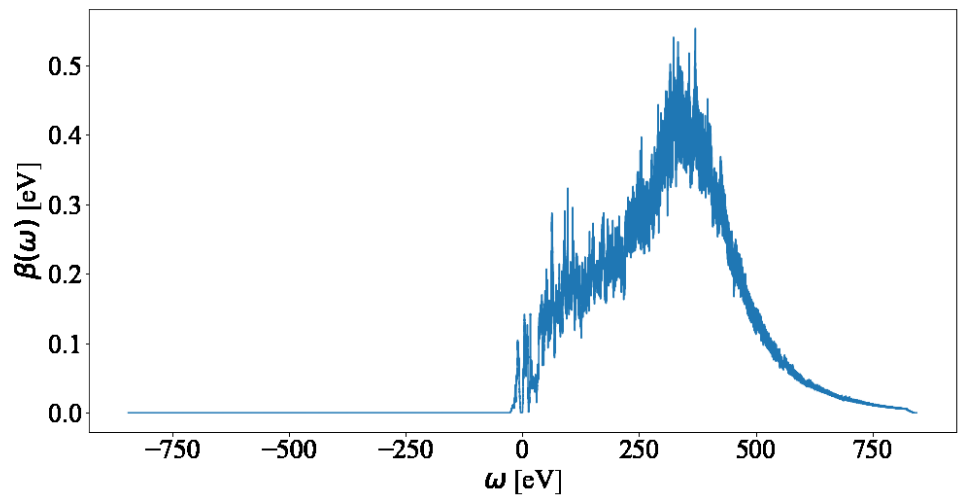}
    \caption{ $\beta_{\text{CBM}}(\omega)$ using a $64\times64\times64$ $\mathbf{k}$- and $\mathbf{q}$-grid with a total of 650 bands.}
    \label{fig:beta}
\end{figure}

\begin{figure}[H]
    \centering
    \includegraphics[width=0.8\columnwidth]{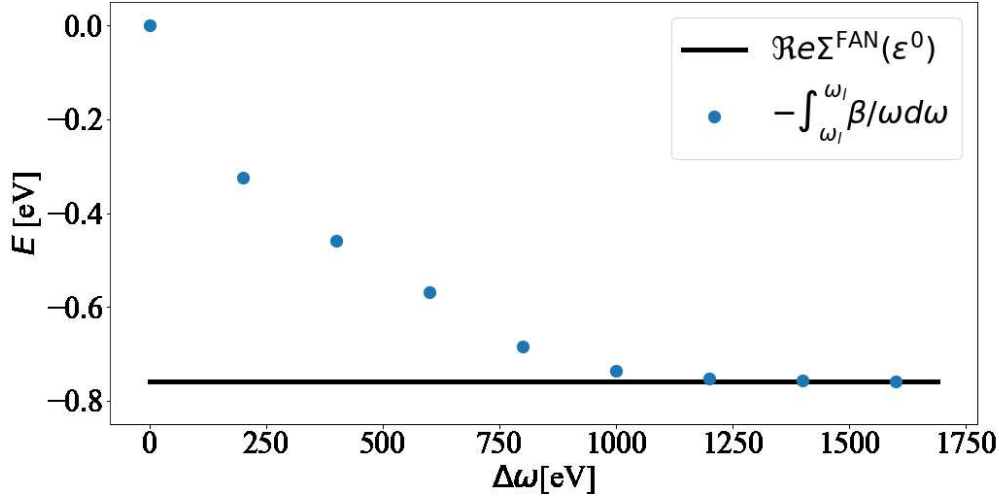}
    \caption{The integration of $\beta_{\text{CBM}}(\omega)/\omega$ ($\beta_{\text{CBM}}(\omega)$ from Fig. \ref{fig:beta}) between $-\omega_l$ and $\omega_l$ ($\Delta \omega)$ converges at a large value of $\Delta \o$, of about 700 eV. The converged value corresponds to the real part of the Fan self-energy at $\omega = \varepsilon^{0}_{\text{CBM}}$ due to the Kramers-Kronig (KK) relation, Eq.~\eqref{eq:KK}. To reduce the computational cost, one can just use KK and avoid the numerical integration of the $\beta_{\text{CBM}}(\o)/\o$ term in Eq.~\eqref{eq:Cumulant}.}
    \label{fig:KK_relationVBM}
\end{figure}

\begin{figure}[H]
    \centering
    \includegraphics[width=0.8\columnwidth]{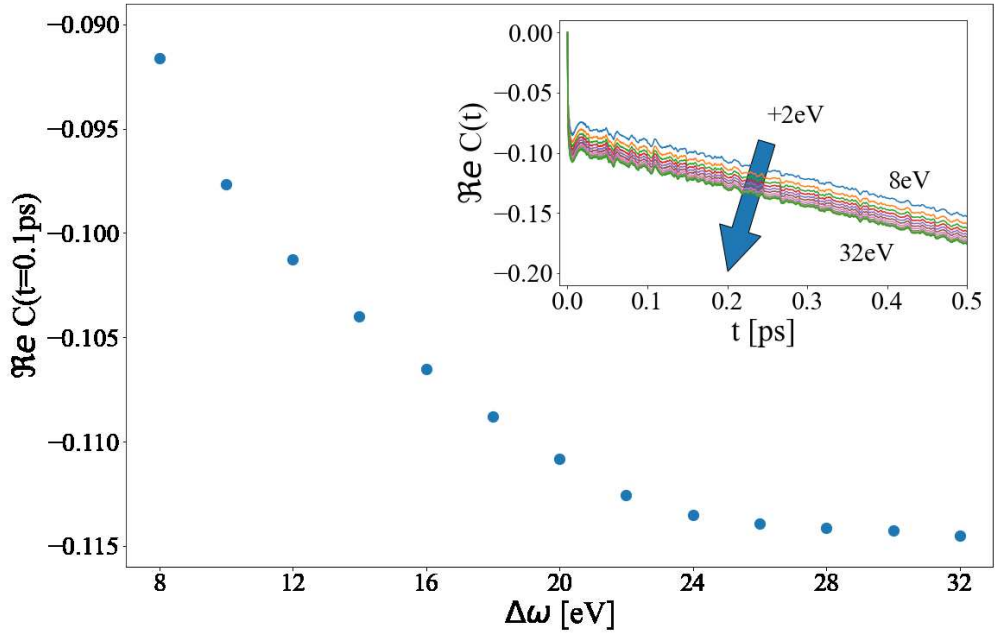}
    \caption{Convergence of $\Re e \tilde{C} (t-0.1 \textrm{ps})$ at the CBM, from Eq. (\ref{eq:Cum_KK}), with respect to the frequency integration range $\Delta \omega$. There is a convergence of 0.001 ps at $\Delta \omega = 22$ eV, which corresponds to 8 bands, and the imaginary part is converged already at $\Delta \omega = 8$ eV. The inset shows $\Re e \tilde{C} (t)$ for different integration ranges. The lines with a decreasing $y$-intercept are calculated for frequency ranges $\Delta \omega$ increasing by 2 eV, from 8 eV and 32 eV, which yields a converging rigid shift of $\Re e \tilde{C} (t)$.}
    \label{fig:FrequencyRange_convergence}
\end{figure}

Converging the $\o$-integral in Eq.~\eqref{eq:Cumulant} requires a very large integration range. In Fig.~\ref{fig:beta} we can see that the imaginary part of the self-energy does not have a small support of $\o$ around $\vare_{n\mathbf{k}}$, but continues to increase up to about 400 eV, and does not go to 0 until values well over 700 eV.
In addition, one of the the cumulant terms, $\int \beta/\o d\o$, has a factor $1/\o$ rather than $1/\o^2$, which makes convergence very slow. In fact, Fig.~\ref{fig:KK_relationVBM} shows that $\Delta \o=$800 eV is needed to converge the integral with an accuracy of 0.01 eV, which corresponds to about 630 bands. This can be avoided by using the Kramers-Kronig (KK) relation,

\be
\Re e\Sigma^\textrm{Fan}_{n\mathbf{k}}(\vare_{n\mathbf{k}})=-P\int^{\infty}_{\infty}\f{\beta_{n\mathbf{k}}(\o)}{\o}d\o.
\label{eq:KK}
\ee

\ni where $P$ indicates the Cauchy principal value. Using this analytical result, the numerical integration of the $1/\o$ term is completely avoided. In this way, we can write

\be
\begin{split}
C_{n\mathbf{k}}(t) & =-i t \Re e\Sigma^\textrm{Fan}_{n\mathbf{k}}(\vare_{n\mathbf{k}}) + P \int_{-\infty}^\infty \b_{n\mathbf{k}}(\o)\f{e^{-i \o t}-1}{\o^2}d\o \\ 
& = -it\Re e\Sigma^\textrm{Fan}_{n\mathbf{k}}(\vare_{n\mathbf{k}}) + \tilde{C}_{n\mathbf{k}}(t),
\end{split}
\label{eq:Cum_KK}
\ee

\ni which is the expression used in our calculations. 
On the other hand,
%the rest of the cumulant, with a $(\textrm{exp}(i\o t)-1)/\o^2$ factor, 
$\tilde{C}(t)$ at $t=0.1$ ps is converged by integrating up to 15 eV (see Fig.~\ref{fig:FrequencyRange_convergence}), which requires the explicit calculation of only 9 bands. For larger values of $t$, the range of energy is even smaller. 

The issue of converging the cumulant has now been reduced to converging the real part of the self-energy in Eq.~\eqref{eq:Cum_KK}, given by Eq.~\ref{eq:Fan}. In principle, this also requires calculating many bands to converge the sum $m$. However, calculating the self-energy is now a standard tool in ABINIT and other first-principles codes, where the sum over bands is usually avoided by using the Sternheimer approximation.
%\textcolor{red}{Joao: It is answered before: 630 bands}\textcolor{blue}{ JP: 630 bands is for the omega integration, which involves the imaginary part and we are now avoiding; now we are looking at the real part. The point is, after avoiding the integration, how many bands do we need? I changed the text a bit.}

%, and

%\be
%G^C(n\mathbf{k},t)=-i\theta(t)e^{-i(\vare_{n\mathbf{k}} + \Re e \Sigma(n\mathbf{k},\vare_{n\mathbf{k}})t)}e^{\tilde{C}(n\mathbf{k},t)},
%\ee

%s\ni 

\section{Sternheimer approximation}\label{sec:Sternheimer}

\begin{figure}[H]
    \centering
    \includegraphics[width=0.8\columnwidth]{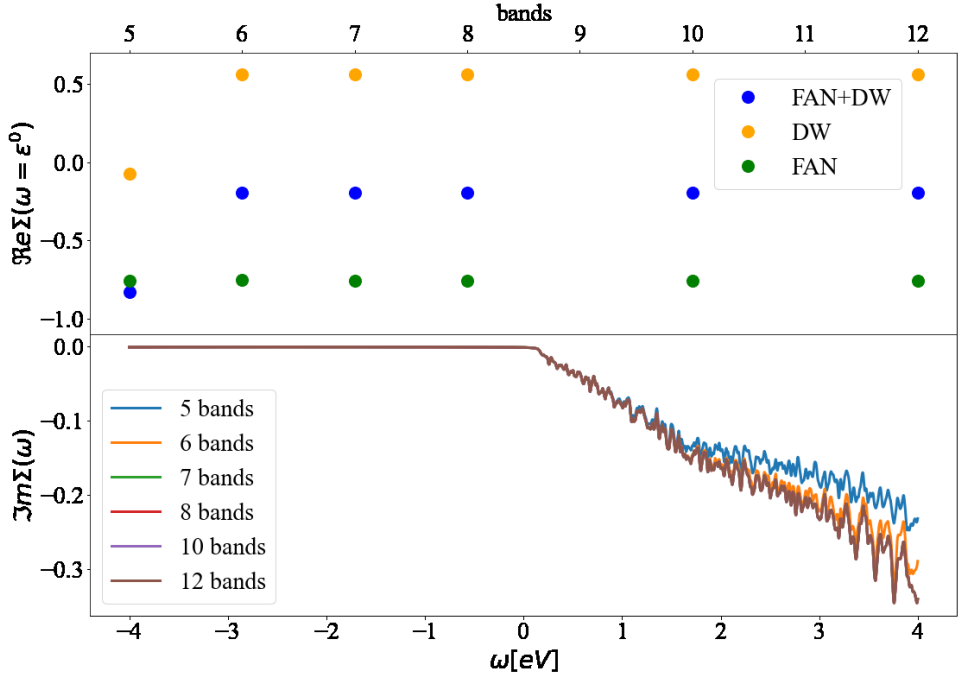}
    \caption{CBM self-energy. Top: Real part at the KS energy as a function of the number of explicit bands $M$. At $M=6$ it is already well converged, explicitly showing that the phonon frequencies play only a small role when energy denominators are large. Bottom: Imaginary part of the self energy, as a function of frequency, for different number of bands. Since the imaginary part is not 0 only if $\omega-\varepsilon_{n'\mathbf{k+q}} \pm \omega_{\mathbf{q}s} = 0$, values agree in the whole range as soon as at least one band is added above 4 eV.}
    %the number of explicit bands. Studying the convergence of Sternheimer equations with respect to bands. The band gap of these calculations is between bands 4 and 5.}
    \label{fig:Sternheimer_convergence}
\end{figure}

Converging the calculation of the e-ph self-energy requires the inclusion of many empty states. This can be circumvented by using Eqs.~(26)-(31) in Ref.\citen{Gonze2010}, in which the sum for $n'>M$ can be replaced by the numerical solution of the Sternheimer equation. Obtaining the self-energy in this way corresponds, for $m>M$ in Eq.~\eqref{eq:Fan}, to setting $\omega=\varepsilon_{n\mathbf{k}}$ (which is precisely what is needed in Eq.~\eqref{eq:KK}) and dropping the phonon frequencies $\omega_{\mathbf{q}s}$ in the denominators.
%This can be avoided by using Eqs.~(26)-(31) in Ref.~\citen{Gonze2010}, in which the sum for $n'>M$ can be replaced by the numerical solution of the Sternheimer equation. This correspond to setting $\omega=\varepsilon_{n\mathbf{k}}$ (which is precisely what is needed in Eq.~\eqref{eq:KK}) and dropping the phonon frequencies $\omega_{\mathbf{q}s}$ in the denominators in Eq.~\eqref{eq:Fan}.
%\textcolor{red}{Joao: I think we can comment this last sentence (not paragraph), these specifics become confusing?} \textcolor{blue}{JP: It's part of the approximation, so if we describe Sternheimer, I think it should be mentioned. To me technical details in the Appendix are fine, unless it's not clear.}
%Notice that $\omega=\varepsilon_{n\mathbf{k}}$ is precisely the case in the left hand side of Eq.~\eqref{eq:KK}. It is also worth pointing out that calculating the energy shift as $\Delta \varepsilon_{n\mathbf{k}}=\Re e \Sigma(n\mathbf{k},\omega=\varepsilon_{n\mathbf{k}})$ is usually a good approximation.
Energy differences $\varepsilon_{n\mathbf{k}}-\varepsilon_{n'\mathbf{k+q}}$ are much larger than $\omega_{\mathbf{q}s}$, even for $m$ not much larger than $n$, so dropping the phonon frequencies has virtually no effect. %of the order of several eV even for small values of $n'$ (and of tens or hundreds of eV for higher bands).
%\textcolor{red}{Joao: Why is this important in this context? I think it is explained in the part commented below or in the next paragraph? Maybe we can comment this part as well?} \textcolor{blue}{JP: It is important to say why neglecting the phonon frequencies is a good approximation. The paragraph below refers to the $\o$ dependence, so it's not the same. Maybe someone else can give their input.}
%In addition, the energy differences $\varepsilon_{n\mathbf{k}}-\varepsilon_{n'\mathbf{k+q}}$ are much larger (of the order of tens of eV or more) than $\omega_{\mathbf{q}s}$, so setting $\omega_{\mathbf{q}s}=0$ in the denominators is fully justified to determine $\Re e \Sigma^\textrm{Fan}(n\mathbf{k},\omega=\varepsilon_{n\mathbf{k}})$, or similarly, $\Delta \varepsilon_{n\mathbf{k}}$.

In Fig.~\ref{fig:Sternheimer_convergence}, we can see that by using the Sternheimer approximation, $\Re e \Sigma^\textrm{Fan}_{n\mathbf{k}}(\omega=\varepsilon_{n\mathbf{k}})$ is converged by summing explicitly over only 6 bands. In the previous section, we obtained that more bands, 9, are needed to converge $\tilde{C}$ in Eq.~\eqref{eq:Cum_KK}, so the method is fully converged by determining only 9 bands of the self-energy. For the VBM, a similar amount of bands are needed. %\textcolor{blue}{I think we can say a smiliar amount of bands are needed}

%In the other cases we considered, a greater amount of bands (?) was also needed to converge $\tilde{C}$ compared to calculating $\Re e \Sigma^\textrm{Fan}(n\mathbf{k},\omega=\varepsilon_{n\mathbf{k}})$ with the Sternheimer approximation. A similar amount of bands (?) were used in the other cases we considered (?). 

Regarding the frequency dependence (dynamical effects) of the self-energy, the Sternheimer approximation implies that it is not included for $m>M$. However, if the real part of the self-energy is evaluated close to $\varepsilon_{n\mathbf{k}}$, such that $|\omega - \varepsilon_{n\mathbf{k}}| \ll |\omega - \varepsilon_{n'\mathbf{k+q}}|$, then this is also a good approximation. % (see Fig.~\ref{fig:ReSigma_bands_conv}). 
%In addition, after using KK, the real part of the self-energy is only needed at $\omega=\xi_{n\mathbf{k}}$. 
For the imaginary part, contributions come from $\omega-\varepsilon_{n'\mathbf{k+q}} \pm \omega_{\mathbf{q}s} = 0$, so as long as $\omega$ takes values slightly lesser than those of the $m>M$ bands, the approximation has no effect, and the spectral function can be accurately determined. Therefore, by using KK, the Sternheimer approximation can be safely used by calculating explicitly the bands up to $m \leq M$, for a small value of $M$.

%As we saw earlier, the $i \omega t$ term in the cumulant requires a very large $\omega$ range and many unoccupied bands to converge, but this is avoided using KK. The other terms of the cumulant require a much narrower range of $\omega$, which is also sufficient to converge $\Re e \Sigma(n\mathbf{k},\omega=\xi_{n\mathbf{k}})$. Therefore, by using KK, the Sternheimer approximation can be safely used by calculating explicitly the bands only up to $n' \leq M$ for a relatively small (moderate?) value of $M$.

% m_dfpt_cgwf.F90 doesn't seem to have an imaginary part

\section{Self-Energy details}\label{sec:SelfEnergy}

The inverse of the Dyson-Migdal equation, eq. \ref{eq: Dyson equation}, is given by
\begin{equation}
    \Sigma_{n \mb{k}} = \frac{1}{G^{(0)}_{n \mb{k} } } - \frac{1}{G_{n \mb{k} } }. 
\end{equation}
There is a distinct decay between $G^0$ and the CE calculated $G$ at large $\omega$, which can be seen in Fig. \ref{fig:G_G0}. The Fourier Transform used for CE method forces $G$ to zero in the frequency numerical limits and this disparity produces a divergence in $\Sigma$ as $G$ tends to zero, Fig. \ref{fig:Sigma_CE_DM}. Nevertheless, close to the KS energy between -2.0 and 2.0 eV, see Fig. \ref{fig:Sigma_CE_DM_zoom}, where $G$ and $G^0$ are close to each other and far from 0 by $\pm$ 0.5 (Fig. \ref{fig:G_G0}), the DM and CE self-energies are within the same range of energies. The real part of the self-energy between the QP line (diagonal solid black line) and the plateau energy line (diagonal dashed black line) seems to begin or continue, depending on the temperature, a descending behaviour. However, reaching to a distance of $\omega_{LO}$ the descent stops and leads to a plateau. This is at variance with polar materials, where at $\omega_{LO}$ the real part of the self-energy diverges.

\begin{figure}[H]
    \centering
    \includegraphics[width=0.8\columnwidth]{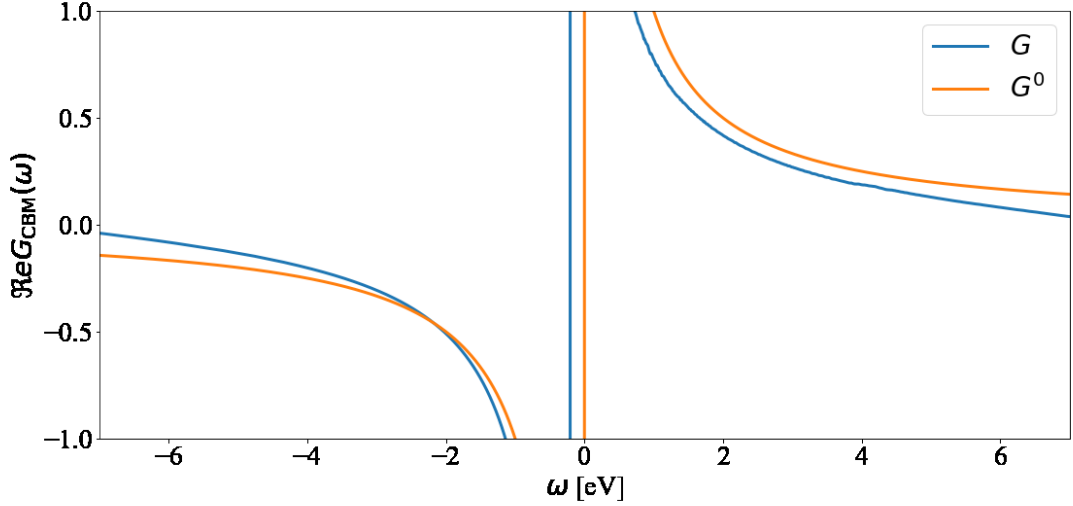}
    \caption{Calculated Green's functions in the frequency domain at CBM. In blue, the total CE Green's function and in orange the non-interacting particle Green's function. The latter has a slower decay to zero (as $1/\omega$) than the former. As the CE G is constructed by Fourier transform, it must be periodic and goes to 0 at the edge of the chosen frequency interval (around $\pm 8$ eV).}
    \label{fig:G_G0}
\end{figure}

\begin{figure}[H]
    \centering
    \includegraphics[width=0.8\columnwidth]{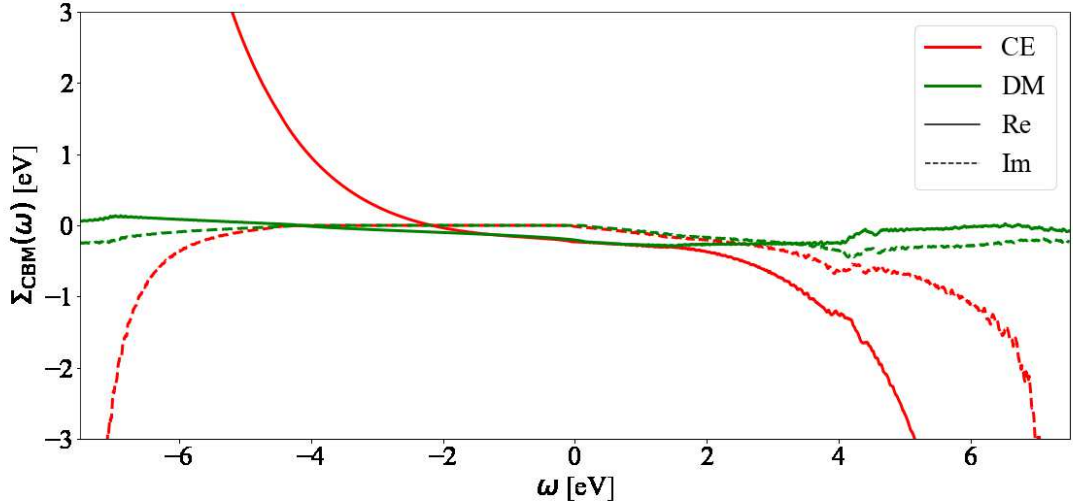}
    \caption{The real (solid) and imaginary (dashed) parts of the self-energy for DM (green) and CE (red) at the CBM. The CE self-energy is calculated as the difference of the inverses of the CE and non-interacting Green's functions, and diverges with increasing $\omega$ due to the distinctive decay (faster than $1/\omega$) of the CE Green's function, Fig. \ref{fig:G_G0}.}
    \label{fig:Sigma_CE_DM}
\end{figure}

\begin{figure}[H]
    \centering
    \includegraphics[width=0.8\columnwidth]{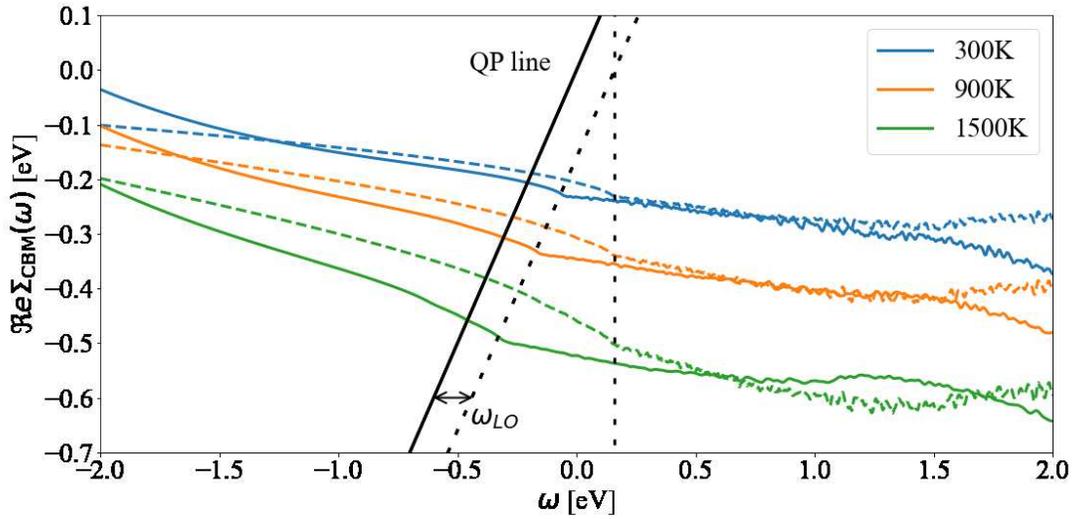}
    \caption{Real part of the self-energy at 300 (blue), 900 (orange), and 1500 (green) K, both for CE (solid) and DM (dashed) calculations at the CBM. The diagonal solid black line shows $\omega = \Re e \Sigma(\omega)$, corresponding to Eq. \eqref{eq:SCQPenergy}, with $\varepsilon^0_{\text{CBM}}$ set to zero, and intersections with $ \Re e \Sigma(\omega)$ yield the QP energy peaks. The diagonal dashed black line is shown at a distance of $\omega_{LO}$ from the QP line. The vertical dashed line is set to $\omega_{LO}$ from $\varepsilon^0_{\text{CBM}}$. Both dashed black lines highlight the position of the satellite plateau structure, which is physical only in the CE case.}
    \label{fig:Sigma_CE_DM_zoom}
\end{figure}

\section{Spectral function and ARPES}

\subsection{Energy values}

Energy values detailed and summarized from Figs. \ref{fig:AwDMandCumulant_Diamond} and  \ref{fig:Aw_DM_CE_VBM} of the main text are compiled in Table \ref{tab:ValuesEnergyTemperature}.

%\begin{widetext}

%\begingroup
%\setlength{\tabcolsep}{2pt} % Default value: 6pt
%\renewcommand{\arraystretch}{1.3} % Default value: 1
%\centering
\begin{table}
 \begin{tabular*}{\textwidth}{@{\extracolsep{\fill}}ccccccc}
 State & \multicolumn{3}{c}{CBM} & \multicolumn{3}{c}{VBM} \\
 Temperature (K) & 300 & 900 & 1500 & 300 & 900 & 1500 \\
 \hline
 $\Re e \Sigma_{n \mb{k}}^{\text{DW}}(\varepsilon^0_{n \mb{k}})$  & 0.659 & 1.148 & 1.779 & 1.974  & 3.440 & 5.333 \\
 $\Re e \Sigma_{n \mb{k}}^{\text{FAN}}(\varepsilon^0_{n \mb{k}})$  & -0.862  & -1.430 & -2.182 & -1.841  & -3.188 & -4.867 \\
 $\Re e \Sigma_{n \mb{k}}(\varepsilon^0_{n \mb{k}})$ & -0.203  & -0.282 & -0.403 & 0.134 & 0.252 & 0.466 \\
 $\Im m \Sigma_{n \mb{k}}(\varepsilon^0_{n \mb{k}})$ & $-1.5 \times 10^{-4}$ & $-3.5 \times 10^{-3}$ & $-9.4 \times 10^{-3}$  &  $-1.1 \times 10^{-3}$ & $-3.9 \times 10^{-2}$ & $-1.1 \times 10^{-1}$ \\
 $\varepsilon^{DM-OMS}$ & -0.203 & -0.282 & -0.403 & 0.134 & 0.252 & 0.466 \\
 $\varepsilon^{DM-Linear}$ & -0.187 & -0.241 & -0.311 & 0.120 & 0.171 & 0.288 \\
 $\varepsilon^{DM-NL}$ & -0.179 & -0.260 & -0.367 & 0.121  & 0.161 & 0.221 \\
 $\varepsilon^{CE-NL}$ & -0.203 & -0.282 & -0.402 & 0.134 & 0.252 & 0.464 \\
 \end{tabular*}
 \caption{Self-energy and energy renormalization values (in eV) at the CBM and VBM for several temperatures. Calculation were done using a \mg{128} $\mb{q}$-grid and $\eta= 5$ meV.}
 \label{tab:ValuesEnergyTemperature}
\end{table}
%\endgroup
%\end{widetext}

\subsection{$\Sp$ and $\Sm$}\label{subsubsec:SigmaPlusSigmaMinus}

In Fig. \ref{fig:Aw_Emission_Absorption}, we separate the contributions $\Sp$ and $\Sm$ of the spectral function at the VBM and set the KS energy at zero. 
$\Sm$ is proportional to $n+1$, so the plateau is visible at low and high temperatures.  Although the self-energy becomes larger with temperature, the spectral function is normalized, and there is little effect in the shape of the plateau. The spectral weight is essentially zero between $-\omega_\mathrm{LO}$ and the peak, indicating that the plateau is indeed given by the contribution of modes with $\o \sim \o_\mathrm{LO}$. As temperature increases, the distance from the QP peak to the plateau is very different from $\o_\mathrm{LO}$.
%which has a frequency (in temperatures units) of more than 1800 K.
%$\Sm$ is proportional to $n+1$, so the plateau is visible at low temperatures, and since the spectral function is normalized, there seems to be little effect. F
For $\Sp$, which is proportional to $n$ instead of $n+1$, there is no plateau at low temperatures. At $T=900$ K, since $+\o_\mathrm{LO}$ is about 1800 K (in temperature units), the plateau should be visible, but it coincides with the main peak, making the peak artificially wide at 900 K. At higher temperatures, the QP peak is shifted to larger values and the plateau acquires more weight, becoming visible.  However, both the overlap of the plateau and main peak, and the varying distance between them, are artificial effects of the DM approach. In CE instead, plateaus (if visible) are at about $-\o_\mathrm{LO}$ and $+\o_\mathrm{LO}$ from the main peak.

For the CBM, Fig. \ref{fig:Aw_Emission_Absorption_CBM}, the behavior is analogous, but the plateau and main peak never merge.
The DM spectral function in Figs. \ref{fig:AwDMandCumulant_Diamond} and \ref{fig:Aw_DM_CE_VBM} uses the full self-energy $\Sigma = \Sm + \Sp$.

%we see that the plateau part does not change with temperature and the QP peak becomes larger (the gap gets smaller). The fact that the spectral weight is essentially 0 between $-\o_\mathrm{LO}$ and the peak, indicates that the plateau is indeed given by the highest phonon branch, which has a frequency (in temperatures units) of more than 1800 K. So, in 

%At 300 K, $\Sm$ reaches the number of states threshold for the nonpolaron that has a binding energy at $-\omega_{LO}$ and stays the same at least until 1500 K. The $\Sp$ is a bit more complex, increasing the temperature allows further states to be populated close to $\omega_{LO}$ and the nonpolaron grows. The broadening of the QP peak at 300 K from $\Sm$ to $\Sp$ corresponds to the intersection of the QP and the range of energy where the nonpolaron is appearing. The QP peak shifts with temperature and at 900 K, the QP energy is at resonance with the nonpolaron, the one we can see close to $\omega_{LO}$ at 1500 K. This explain the complex structure of the spectral function in Fig. \ref{fig:Aw_DM_CE_VBM}.

\begin{figure}[H]
    \centering
    \includegraphics[width=0.8\columnwidth]{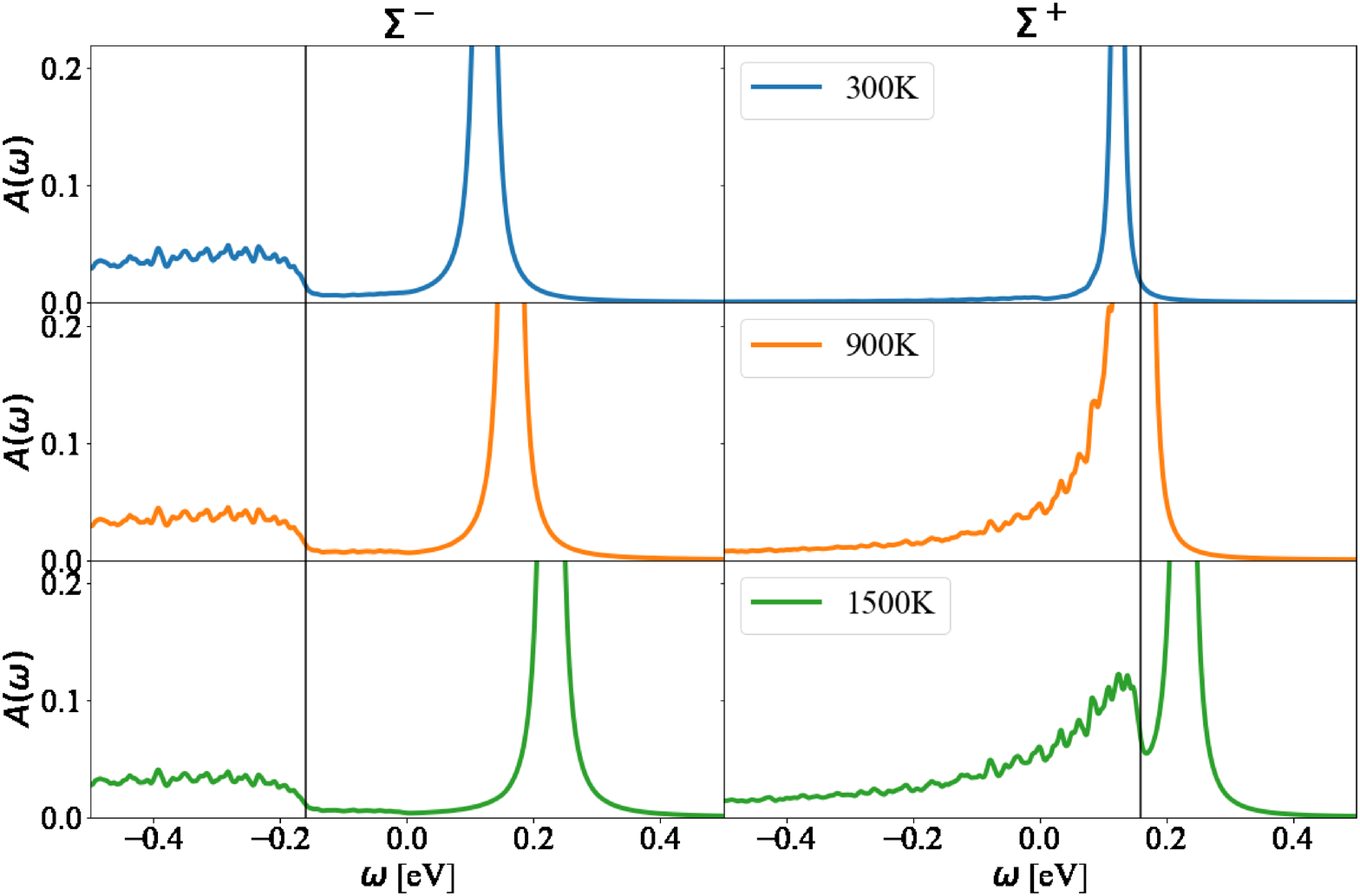}
    \caption{The DM spectral function at the VBM splitted into $\Sp$ and $\Sm$ for T=300, 900, 1500 K, calculated with a \mg{128} $\mb{q}$-mesh and $\eta=$ 5 meV. The vertical black lines in the figures show $-\omega_{LO}$ on the left and $+\omega_{LO}$ on the right.}
    \label{fig:Aw_Emission_Absorption}
\end{figure}

%For the CBM, Fig. \ref{fig:Aw_Emission_Absorption_CBM}, we see the opposite, the nonpolaron reaches its threshold at 300 K in the $\Sp$ and, contrary to the VBM, the QP peak is not inside the energy range of the nonpolaron. Therefore, the QP peak has no extra broadening due to the nonpolaron. The nonpolaron is not yet present at 300 K for $\Sm$, but increasing the temperature, there are states starting at $\omega_{LO}$ that are populated to create the nonpolaron.

\begin{figure}[H]
    \centering
    \includegraphics[width=0.8\columnwidth]{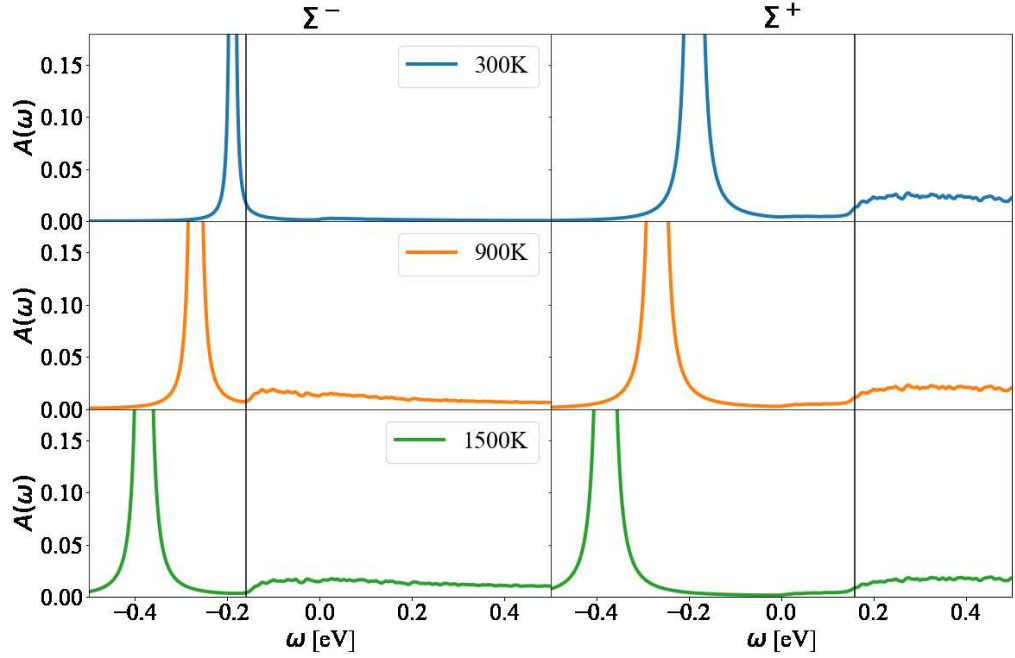}
    \caption{The DM spectral function at CBM splitted into $\Sp$ and $\Sm$ for T=300, 900, 1500 K, calculated with a \mg{128} $\mb{q}$-mesh and $\eta=$ 5 meV. The vertical black lines in the figures show $-\omega_{LO}$ on the left and $+\omega_{LO}$ on the right.}
    \label{fig:Aw_Emission_Absorption_CBM}
\end{figure}

\subsection{Scattering matrix elements}

The e-ph scattering matrix elements are obtained from the derivative of the KS potential. Although quadrupoles contribute to the KS potential around $\Gamma$, their contribution to the electron-phonon matrix elements is negligible, as can be observed in Fig.~\ref{fig:el-phmatrixelements}. $\left| g_{CBM} (\mb{q}) \right|$ is the sum of the matrix elements over all bands and phonon modes calculated at the CBM.

\begin{figure}[H]
    \centering
    \includegraphics[width=0.8\columnwidth]{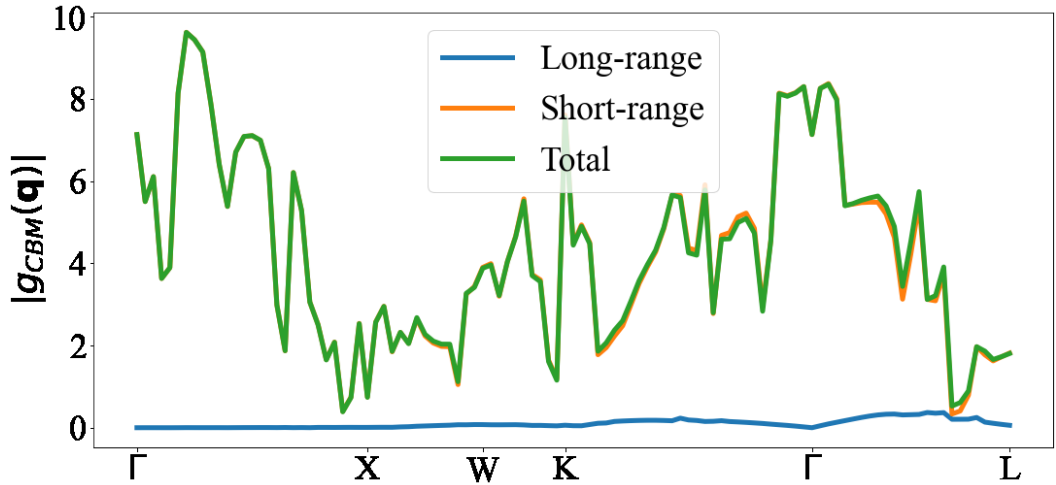}
    \caption{Modulus of electron-phonon scattering matrix elements, $g_{nm \mb{k}}^{j \mb{q}}$, evaluated at $n \mb{k} = \text{CBM}$ and summed over the $m$ bands and $j$ phonon modes. These elements are pictured in a $\mb{q}$-path following high-symmetric crystal points in the reciprocal lattice. There is a split between long-range (blue) and short-range fields (orange), together with their sum (green). The long-range contribution to the electron-phonon scattering matrix elements, and, thus, to the nonpolaron plateau is negligible.  }
    \label{fig:el-phmatrixelements}
\end{figure}

%An electronic carrier at the CBM $\mb{k}$-point interacts with the state $\mb{k} + \mb{q}$ by absorbing or emitting a phonon with wave-vector $\mb{q}$. The reciprocal space position of the carrier in Fig. \ref{fig:el-phmatrixelements} is at $\Gamma$ and the other states are viewed as displaced from the CBM by the vector $\mb{q}$. The states close to $\Gamma$ have stronger interaction with the state $\Gamma$ itself (CBM). There is a high peak between $\Gamma$ and X, which corresponds to the degeneracy of bands close to the CBM at X shown in Fig. \ref{fig:ARPES_CBM}. While the electronic $\Gamma$ $\mb{k}$-point  is situated close to the X $\mb{k} + \mb{q}$-point with zero strength and surrounded by low values, showing even further the inconsequential effect of long-range fields. 

\subsection{ARPES}

Taking into account just the ARPES theoretical picture, Fig. \ref{fig:ARPES}, it is difficult to discern changes between DM and CE approaches at 300 K, Fig. \ref{fig:ARPES300K}. As the temperature is increased to 900 K, Fig. \ref{fig:ARPES900K}, or 1500 K, Fig. \ref{fig:ARPES1500K}, one can notice a shift and broadening of the bands, in particular at the CBM between $\Gamma$ and $X$ points and the VBM at $\Gamma$ point. The horizontal lines visible at 8 eV and (less visible) at -3 eV, corresponding to a very small peak in the spectral function, are due to the high density of states at the band extrema at L.

\begin{figure}[h!]
     \centering
     \begin{subfigure}[b]{\columnwidth}
         \centering
         \includegraphics[width=10cm,height=6.5cm]{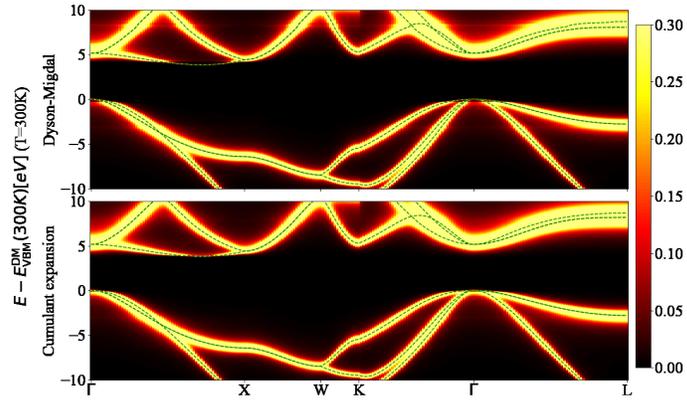}
         \caption{$T=300K$}
         \label{fig:ARPES300K}
     \end{subfigure}
     \begin{subfigure}[b]{\columnwidth}
         \centering
         \includegraphics[width=10cm,height=6.5cm]{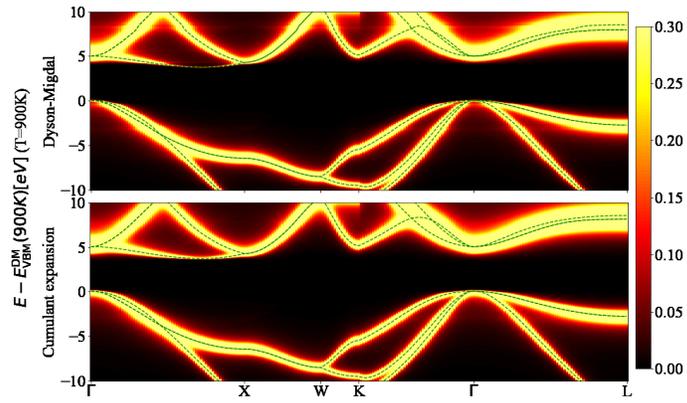}
         \caption{$T=900K$}
         \label{fig:ARPES900K}
     \end{subfigure}
     \begin{subfigure}[b]{\columnwidth}
         \centering
         \includegraphics[width=10cm,height=6.5cm]{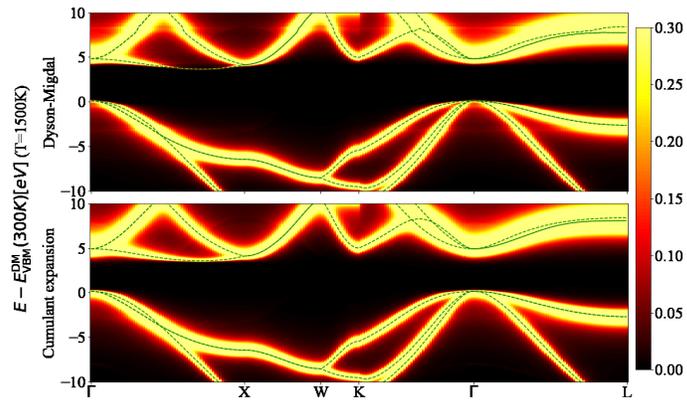}
         \caption{$T=1500K$}
         \label{fig:ARPES1500K}
     \end{subfigure}
        \caption{Spectral function at 300 K (top), 900 K (middle), and 1500 K (bottom). The calculations were performed using DM (top in each figure) and CE (bottom in each figure) with a $N=128$ $\mb{q}$-grid. All values were displaced in relation to the VBM at 300 K. To be able to observe the presence of the nonpolaron signature near VBM and CBM, the intensity scale of the density of states (colormap) was limited to 0.3.}
        \label{fig:ARPES}
\end{figure}

 \end{@twocolumnfalse}

\end{document}